\documentstyle[12pt,fleqn,rotate,epsfig]{article}
\newcommand{\be}{\begin{equation}}
\newcommand{\ee}{\end{equation}}
\newcommand{\bea}{\begin{eqnarray}}
\newcommand{\eea}{\end{eqnarray}}

\begin{document}
\title{Antikaon production  in $A+A$ collisions at SIS
energies within an off-shell G-matrix approach\footnote{supported
by DFG and GSI Darmstadt}}
\author{W. Cassing$^1$, L. Tol\'os$^2$, E. L. Bratkovskaya$^3$,
A. Ramos$^2$\\ \\
$^1$Institut f\"ur Theoretische Physik, Universit\"at Giessen\\
D-35392 Giessen, Germany \\
$^2$Departament d'Estructura i Constituents de la Mat\`eria,\\
Universitat de Barcelona, \\ Diagonal 647,
08028 Barcelona, Spain \\
$^3$Institut f\"ur Theoretische Physik, Universit\"at Frankfurt/M\\
D-60054 Frankfurt, Germany }
\date{  }
\maketitle
\begin{abstract}
The production and propagation of antikaons -- described by
dynamical spectral functions $A_h(X,\vec{P},M^2)$  as evaluated
from a coupled channel $G$-matrix approach -- is studied for
nucleus-nucleus collisions at SIS  energies in comparison to the
conventional quasi-particle limit and the available experimental
data using off-shell transport theory. We find that the $K^-$
spectra for $^{12}C + ^{12}C$ and $^{58}Ni + ^{58}Ni$ at 1.8
A$\cdot$GeV remain  underestimated in the $G$-matrix approach as
in the on-shell quasi-particle approximation whereas the
preliminary spectra for $Au + Au$ at 1.5 A$\cdot$GeV are well
described in both limits. This also holds approximately for the
$K^-$ rapidity distributions in semi-central collisions of $Ni+Ni$
at 1.93 A$\cdot$GeV. However, in all limits considered  there is
no convincing description of all spectra simultaneously. Our
off-shell transport calculations, furthermore, demonstrate that
the strongest in-medium effects should be found for low antikaon
momenta in the center-of-mass frame, since the deceleration of the
antikaons in the attractive Coulomb and nuclear potentials and the
propagation to the on-shell mass induces a net shift and squeezing
of the $K^-$ spectra to the low momentum regime.
\end{abstract}


%


Relativistic heavy-ion collisions; $\bar{K}N$ interaction

\newpage
\section{Introduction}
\label{introduction} An open problem of todays strong interaction
physics is the dynamical generation of hadron masses from the
quark and gluon fields, that are the elementary fields of
Quantum-Chromo-Dynamics (QCD). Whereas QCD lattice calculations
nowadays give quite reliable results for the masses of the $0^-$
and $1^-$ meson octet states - except for the light pions - as
well for the baryon octet and decuplet states, the masses and
properties of these states are not well known at finite quark
chemical potential $\mu_q$ or finite baryon density $\rho_B$. This
is essentially due to the problem of calculating the fermion
determinant at finite $\mu_q$. Though some recent progress has
been made for QCD lattice calculations at finite baryon density
\cite{xxx,Allton}, the properties of hadrons in a baryonic
environment cannot be addressed directly so far by such {\it
ab-initio} calculations. One thus presently has to rely on
effective hadronic Lagrangians and to compute the hadron
properties at finite density $\rho_B$ and/or temperature $T$ by
'dressing' the vacuum states in the medium e.g. within $G$-matrix
theory. An alternative way is to employ effective chiral
Lagrangians and to extract the leading order effects for the
hadron modifications in the medium.

As demonstrated in the pioneering work of Kaplan and Nelson
\cite{Kaplan,nelson} within a chiral approach kaons and
antikaons couple attractively to the
scalar nucleon density with a strength proportional to the $KN-
\Sigma$ constant,
\begin{equation} \label{sigmat}
\Sigma_{KN} = \frac{1}{2} (m^0_u + m^0_s) \ <N|\bar{u}u +
\bar{s}s|N>,
\end{equation}
which may vary  from 270--450 MeV (cf. the discussion in Ref.
\cite{Schaffner}). QCD lattice calculations here provide further
information \cite{latnew}. In Eq. (\ref{sigmat}) $m^0_u$ and $m^0_s$
denote the bare masses for the light $u$- and strange $s$-quark
while the expression in the brackets is the expectation value of
the scalar light and strange quark condensate for the nucleon
\cite{CB99}. Furthermore, a vector coupling to the quark 4-current
-- for vanishing spatial components -- leads to a repulsive
potential term for the kaons; this (Weinberg-Tomozawa) term is
attractive for the antikaons.

In relativistic mean-field models the dispersion relation for
kaons and antikaons in the nuclear medium -- for low momenta --
can be written as \cite{EX1,EX2,EX3}
\begin{equation} \label{nelson1}
\omega_{K^\pm}(\rho_N,{\bf p}) = \pm \frac{3}{8}
\frac{\rho_N}{f_\pi^2}  + \left[ {\bf p}^2 + m_K^2 \left( 1 -
\frac{\Sigma_{KN}}{f_\pi^2 m_K^2} \rho_s + \left( \frac{3\rho_N}{8
f_\pi^2 m_K} \right)^2 \right) \right]^{1/2} .
\end{equation}
In Eq. (\ref{nelson1}) $m_K$ denotes the bare kaon mass,
$f_\pi \approx$ 93 MeV is the pion decay constant, while $\rho_s$
and $\rho_N$ stand for the scalar and vector nucleon densities,
respectively.  Note that, when extrapolating Eq. (\ref{nelson1}) to
3$\rho_0$ and above, the antikaon mass becomes very light. Thus
antikaon condensates might occur at high baryon density which,
furthermore, are of interest in the astrophysical context
\cite{GB1,muto,sahu01,EX4}.

Studies on $K^\pm$ production from nucleus-nucleus collisions
at SIS energies of 1-- 2 A$\cdot$ GeV have shown that in-medium
properties of kaons are seen in the collective flow pattern of
$K^+$ mesons, both in-plane and out-of-plane, as well as in the
abundancy and spectra of antikaons
\cite{CB99,cmko,lix,Li2001,K5,K6}. Thus in-medium modifications of
the mesons have become a topic of substantial interest in the last
decade triggered in part by the early suggestion of Brown and Rho
\cite{BR}, that the modifications of hadron masses should scale
with the scalar quark condensate $<q\bar{q}>$ at finite baryon
density.

The actual kaon and antikaon self energies (or potentials) are
quite a matter of debate and depend on the many-body resummation
scheme adopted. Especially for the antikaons  the
momentum-dependence of their self energies is widely unknown since
most Lagrangian models restrict to $s$-wave interactions or only
include additional $p$-waves \cite{Lutz021,Boris,jido}. Here only
limited information is available from a dispersion analysis in
Ref. \cite{sibkaon}.

There have been attempts to extract the antikaon-nucleus potential
from the analysis of kaonic-atom data and some solutions favor
very strongly attractive potentials of the order of -200 MeV at
normal nuclear matter density $\rho_0$ \cite{FGB94,Gal}. However,
more recent self-consistent calculations based on a chiral
Lagrangian \cite{Lutz98,Oset00} or meson-exchange potentials
\cite{Laura} only predict moderate attractive depths of -50 to -80
MeV at density $\rho_0$.  In addition, studies of kaonic atoms
using the chiral ${\bar K}N$ amplitudes of Ref.~\cite{oset98} show
that it is indeed possible to find a reasonable reproduction of
the data with a relatively shallow antikaon-nucleus potential
\cite{satoru00,Ramos}, albeit adding an additional moderate
phenomenological part \cite{baca}. This has been corroborated by a
calculation in Ref. \cite{cieply01}, where a good fit to both
scattering $K^- p$ data and kaonic-atom data only required to
modify slightly the parameters of the chiral meson-baryon
interaction model of Ref.~\cite{weise}.
Since it is clear that kaonic atom data do not suitably constrain
the antikaon-nucleus potential at normal nuclear matter density, a
recent work \cite{Carmen} explored the possibility of
distinguishing between deep or shallow potentials
\cite{FGB94,Oset00,baca} by means of nuclear scattering of low
energy $K^-$ produced from $\Phi$ decay at Da$\Phi$ne.

In fact, the antikaon-nucleon amplitude in the isospin channel
$I=0$ is dominated by the $\Lambda(1405)$ resonant structure,
which in free space appears only 27 MeV below the ${\bar K}N$
threshold. This resonance is generated dynamically from a coupled
channel $T$-matrix scattering equation using a suitable
meson-baryon potential. The coupling between the ${\bar K}N$ and
$\pi Y$ ($Y=\Lambda,\Sigma$) channels is essential to get the
right dynamical behavior in free space. Correspondingly, the
in-medium properties of the $\Lambda(1405)$, such as its pole
position and its width, which in turn influence strongly the
behavior of the antikaon-nucleus optical potential, are very
sensitive to the many-body treatment of the medium effects.
Previous works have shown, for instance, that a self-consistent
treatment of the $\bar{K}$ self energy has a strong influence on
the scattering amplitudes \cite{Lutz98,Oset00,Laura,Schaffner2}
and, consequently, on the in-medium properties of the antikaon.
Moreover, the incorporation of the pion with its medium modified
properties also proved to be an important aspect
\cite{Oset00,lauran}, although most works until now have ignored
it. As pointed out in Ref. \cite{Lutz02} also the properties of
the $\Sigma^*(1385)$ might be essential for in-medium transitions
rates at finite relative momentum since the $\Sigma^*(1385)$ is
the strange analogue to the $\Delta(1232)$.

Furthermore, a sizeable progress in the understanding of hadron
off-shell propagation in phase-space -- even for complex
configurations -- has been obtained in the last years. In Refs.
\cite{Cass99,Cass99b} a semiclassical transport approach has been
developed on the basis of the Kadanoff-Baym equations that
includes the propagation of hadrons with dynamical spectral
functions.  This approach has been examined for nucleus-nucleus
collisions at GANIL \cite{Cass99}, SIS and AGS energies
\cite{Cass99b} as well as for equilibration phenomena in related
infinite nuclear matter problems \cite{Cass99c}. Formally, the
input for the off-shell transport approach with respect to the
complex hadron self energies and the various channel transition
probabilities should be provided by coupled-channel $G$-matrix
calculations. Though being far from complete, the present work
makes a major step in this direction by incorporating the
off-shell information from $\bar{K} N \leftrightarrow Y\pi$
transitions.

The paper is organized as follows: In Section 2 we will briefly review
the generalized transport equations on the basis of the Kadanoff-Baym
equations \cite{kb62} and present the dynamical equations of motion
for 'test-particle' propagation in 8-dimensional phase space in
comparison to the traditional on-shell equations of motion. The input
for this approach relates to self-consistent self energies for the
hadrons which here are taken from the coupled-channel $G$-matrix
approach following Ref.  \cite{lauran} (Section 3). The actual
description of in-medium kaon and antikaon production in $NN$ and $\pi
N$ collisions - in line with the $\bar{K}$ spectral function from
the $G$-matrix approach -- is described in Section 4. The explicit
transport calculations for nucleus-nucleus collisions in the SIS
energy regime are presented in Section 5 in comparison to the
experimental data available. A summary and discussion of open problems
concludes this study in Section 6.

\section{Description of the off-shell transport approach}
Whereas the equilibration of strongly interacting quantum systems
has been studied on the basis of the Kadanoff-Baym equations
\cite{kb62} for infinite nuclear matter problems quite some time
ago \cite{Dani} the question how to propagate 'broad resonances'
in an inhomogenuous medium out-of equilibrium (as encountered in
relativistic nucleus-nucleus collisions), has been addressed and
solved only in the last years
\cite{Cass99,Cass99b,Cass99c,Leupold,Knoll1,Knoll2,Knoll3,Knoll4}.
In the latter works a semi-classical off-shell transport approach
has been derived from the Wigner transformed Kadanoff-Baym
equations in the limit of first order gradients in phase space.
The approach by Ivanov et al. in Refs.
\cite{Knoll1,Knoll2,Knoll3,Knoll4} differs from those in Refs.
\cite{Cass99,Cass99b,Cass99c,Leupold} in the treatment of the
'rearrangement term', which is kept in all orders in Refs.
\cite{Knoll1,Knoll2,Knoll3,Knoll4}, but only employs first order
phase-space gradients  in Refs.
\cite{Cass99,Cass99b,Cass99c,Leupold} as suggested
 by Botermans and Malfliet already in 1990 \cite{Bot}. The latter limit
allows to perform a test-particle solution to the problem, which
is adequate for present transport approaches, whereas the
transport equations from Ivanov et al.
\cite{Knoll1,Knoll2,Knoll3,Knoll4} might only be solved on a
8-dim. grid in phase space which is very unconvenient for actual
applications. A recent overview on the different approaches has
been given in Ref. \cite{Knoll4}.

We here follow the off-shell approach developed in Refs.
\cite{Cass99,Cass99b,Cass99c}. For the actual details we refer the
reader to the original articles or independently to Ref.
\cite{Leupold} for a nonrelativistic formulation. We now
concentrate on those results that are important for our present
study. First of all, the transport limit provides an algebraic
result for the hadron spectral function  \bea A_{XP} \; = \ \frac{
\Gamma_{XP} } {( \, P^2 \, - \, M_{0}^{2} \, - \, Re
\Sigma^{ret}_{XP} )^{2} \: + \: \Gamma_{XP}^{2}/4} \; ,
\label{spectral} \eea which holds for all approaches
\cite{Cass99,Cass99b,Cass99c,Leupold,Knoll1,Knoll2,Knoll3,Knoll4}.
In Eq. (\ref{spectral}) $M_0$ denotes the bare mass pole,
$\Gamma_{XP} = -2 Im \Sigma^{ret}_{X,P}$ while
$\Sigma^{ret}_{X,P}$ is the retarded self energy of the hadron
which in general depends on space-time $X$ and the four-momentum
$P$ in a hadronic environment.

\subsection{Testparticle representation}

In order to obtain an approximate solution to the transport
equation ((16) in Ref. \cite{Cass99b}) we use a testparticle
ansatz for the Green function $S^{<}_{XP}$, i.e. for the Wigner
transform of
\begin{equation}
i \, S^{<}_{xy}  :=  \; < \, \Phi^{\dagger}(y) \; \Phi(x) \, > \, ,
\end{equation}
with $X=(x+y)/2$, where $\Phi(x), \Phi^{\dagger}(y)$ denote the
hadron field operators at space-time position $x$ or $y$. More
specifically, we rewrite $iS^{<}_{XP}$ in terms of the real and
positive semidefinite quantity
\bea F_{XP} \; = A_{XP} N_{XP} = \; i \, S^{<}_{XP} \; \sim \;
\sum_{i=1}^{N} \; \delta^{(3)} ({\vec{X}} \, - \, {\vec{X}}_i (t))
\; \; \delta^{(3)} ({\vec{P}} \, - \, {\vec{P}}_i (t)) \; \;
\delta(P_0 - \, \epsilon_i(t)) \: , \label{testparticle} \eea
where the spectral function $A_{XP}$ [Eq.~(\ref{spectral})] is separated
from the number density function $N_{XP}$. In the most general
case (where the self energies depend on four-momentum $P$, time
$t$ and the spatial coordinates $\vec{X}$) the equations of motion
for the testparticles  read \cite{Cass99b}
\bea \label{eomr}
\frac{d {\vec X}_i}{dt} \! & = & \! \phantom{- } \frac{1}{1 -
C_{(i)}} \, \frac{1}{2 \epsilon_i} \: \left[ \, 2 \, {\vec P}_i \,
+ \, {\vec \nabla}_{P_i} \, Re \Sigma^{ret}_{(i)} \, + \, \frac{
\epsilon_i^2 - {\vec P}_i^2 - M_0^2 - Re
\Sigma^{ret}_{(i)}}{\Gamma_{(i)}} \: {\vec \nabla}_{P_i} \,
\Gamma_{(i)} \: \right],
\\[0.3cm]
\label{eomp}
\frac{d {\vec P}_i}{d t} \! & = & \!
- \frac{1}{1-C_{(i)}} \,
\frac{1}{2 \epsilon_{i}} \:
\left[ {\vec \nabla}_{X_i} \, Re \Sigma^{ret}_i
\: + \: \frac{\epsilon_i^2 - {\vec P}_i^2 - M_0^{2}
- Re \Sigma^{ret}_{(i)}}{\Gamma_{(i)}}
\: {\vec \nabla}_{X_i} \, \Gamma_{(i)} \:
\right],
\\[0.3cm]
\label{eome}
\frac{d \epsilon_i}{d t} \! \setlength{\mathindent}{-0.5cm}
& = & \!
\phantom{- }
\frac{1}{1 - C_{(i)}} \,
\frac{1}{2 \epsilon_i} \:
\left[ \frac{\partial Re \Sigma^{ret}_{(i)}}{\partial t}
\: + \: \frac{\epsilon_i^2 - {\vec P}_i^2 - M_0^{2}
- Re \Sigma^{ret}_{(i)}}{\Gamma_{(i)}}
\: \frac{\partial \Gamma_{(i)}}{\partial t}
\right],
\eea\\
where the notation $F_{(i)}$ implies that the function is taken at
the coordinates of the testparticle, i.e.
$F_{(i)} \equiv F(t,\vec{X}_{i}(t),\vec{P}_{i}(t),\epsilon_{i}(t))$.

In Eqs.(\ref{eomr})-(\ref{eome}) the common multiplication factor $(1-C_{(i)})^{-1}$
contains the energy derivatives of the retarded self energy
\bea
\label{correc}
C_{(i)} \: = \:
\frac{1}{2 \epsilon_i}
\left[
\frac{\partial}{\partial \epsilon_i} \, Re \Sigma^{ret}_{(i)} \: + \:
\frac{\epsilon_i^2 - {\vec P}_i^2 - M_0^2
- Re \Sigma^{ret}_{(i)}}{\Gamma_{(i)}}
\: \frac{\partial }{\partial \epsilon_i} \,
\Gamma_{(i)}
\right] \:
\eea\\
which  yields a shift of the system time $t$ to the 'eigentime' of
particle $i$ defined by $\tilde{t}_{i} = t /(1-C_{(i)})$. As in
Refs. \cite{Cass99b,Cass99c} we will assume $C_{(i)} =0$
furtheron. We mention, that the approximation $C_{(i)} =0$  is
also employed in the more standard transport models that operate
with on-shell quasi-particles. In the limiting case of particles
with vanishing gradients of the width $\Gamma_{XP}$ the equations
of motion (\ref{eomr}) - (\ref{eome}) reduce to the well-known
transport equations of the quasi-particle picture.

Furthermore, following Ref. \cite{Cass99} we take
$M^{2} = P^2 - Re \Sigma^{ret}$ as
an independent variable instead of $P_0$, which then fixes the energy
(for given $\vec{P}$ and $M^{2}$) to
\bea
P_{0}^{2} \; = \; \vec{P}^{2} \: + \: M^{2} \: + \:
Re \Sigma_{X\vec{P}M^2}^{ret} \, .
\label{energyfix}
\eea
Eq. (\ref{eome}) then leads to
\bea
\label{eomm}
\frac{dM_i^2}{dt} \; = \;
\frac{M_i^2 - M_0^2}{\Gamma_{(i)}} \;
\frac{d \Gamma_{(i)}}{dt}
\eea
for the time evolution of the test-particle $i$ in the invariant
mass squared as derived in Ref. \cite{Cass99b}. It is worth noting
that Eq. (\ref{eomm}) (for $\Gamma_{(i)} \neq 0$) is equivalent to
\begin{equation} \label{ofinal}
\frac{d}{dt} \left( \frac{M_i^2 - M_0^2}{\Gamma_{(i)}} \right) = 0 ,
\end{equation}
which states that the ratio of $\Delta M_i^2 = M_i^2 - M_0^2$ to the actual width
$\Gamma_{(i)}$ is a constant in time.

\subsection{Collision terms}

The collision term of the Kadanoff-Baym equation in first order gradient expansion
reads as \cite{Cass99b}
$$
I_{coll}(X,\vec{P},M^2) = Tr_2 Tr_3 Tr_4 A(X,{\vec P},M^2)
A(X,{\vec P}_2, M_2 ^2) A(X,{\vec P}_3, M_3 ^2)
A(X,{\vec P}_4, M_4 ^2)
$$
$$
|G(({\vec P},M^2) + ({\vec P}_2,M_2^2)
\rightarrow ({\vec P}_3,M_3^2) + ({\vec P}_4,M_4^2))|_{{\cal A,S}}^2
\; \; \delta^{(4)}({P} + {P}_2 - {P}_3 - {P}_4)
$$
\be
\label{Icoll}
[ \, N_{X{\vec P}_3 M_3^2} \, N_{X {\vec P}_4 M_4^2} \,
{\bar f}_{X {\vec P} M^2} \, {\bar f}_{X {\vec P}_2 M_2^2}
\: - \: N_{X{\vec P} M^2} \, N_{X {\vec P}_2 M_2^2} \,
{\bar f}_{X {\vec P}_3 M_3^2} \,
{\bar f}_{X {\vec P}_4 M_4^2} \, ]
\ee
with
\be
\label{pauli} {\bar f}_{X {\vec P} M^2} = 1 + \eta \, N_{X {\vec P} M^2} \ee
and $\eta = \pm 1$ for bosons/fermions, respectively. The indices
${\cal A,S}$ stand for the antisymmetric/symmetric matrix element
of the in-medium scattering amplitude $G$ in case of
fermions/bosons. In Eq. (\ref{Icoll}) the trace over particles
2,3,4 reads explicitly for fermions
\be
\label{trace}
Tr_2 = \sum_{\sigma_2, \tau_2} \frac{1}{(2 \pi)^4}
\int d^3 P_2 \frac{d M^2_2}{2 \sqrt{\vec{P}^2_2+M^2_2}},
\ee
where $\sigma_2, \tau_2$ denote the spin and isospin of particle 2.
In case of bosons we have
\be
\label{trace2} Tr_2 = \sum_{\sigma_2, \tau_2} \frac{1}{(2 \pi)^4}
\int d^3 P_2 \frac{d P_{0,2}^2}{2}, \ee since the spectral
function $A_B$ is normalized as
\be
\label{sb} \int \frac{d P_0^2}{4 \pi} A_B(X,P) = 1 \ee
whereas for fermions we get
\be
\label{sb1}
\int \frac{d P_0}{2 \pi} A_F(X,P) = 1.
\ee
Here the spectral function $A_F$ in case of fermions in
Eq.~(\ref{Icoll}) is obtained
by considering only particles of positive energy and assuming the spectral
function to be identical for spin 'up' and 'down' states.

  We note that the extension of Eq. (\ref{Icoll}) to inelastic
scattering processes (e.g. $\bar{K}N \rightarrow Y \pi$) or ($\pi
Y \rightarrow \bar{K} N$ etc.) is straightforward when exchanging
the elastic transition amplitude $G$ by the corresponding
inelastic one and taking care of Pauli-blocking or
Bose-enhancement for the particles in the final state. For bosons
we will neglect a Bose-enhancement factor throughout this work
since their actual phase-space density is small for the systems of
interest.

 Thus the transport
approach and  the particle spectral functions are fully determined once
 the in-medium transition amplitudes $G$ are
known {\it in their full off-shell dependence}. For the
transitions $NN \rightarrow NN$ and $N \Delta \leftrightarrow NN$
we employ the same approximations as in Ref. \cite{Cass99b}
whereas for the strangeness sector $\bar{K} N \leftrightarrow
\bar{K} N$ and $\bar{K} N \leftrightarrow \pi Y$ we will use the
off-shell transition rates as determined from the coupled channel
$G$-matrix approach to be described below.

\section{The coupled-channel $G$-matrix approach}

In Ref.~\cite{Laura}, the effective $\bar{K} N$ interaction in the
nuclear medium ($G$-matrix) at temperature $T=0$ was derived from
a meson-baryon bare interaction built in the meson exchange
framework \cite{holinde90}. As the bare interaction permits
transitions from the $\bar {K} N$ channel to the $\pi \Sigma$ and
$\pi \Lambda$ ones, all having strangeness $S=-1$, one is
confronted with a coupled-channel problem.  Working in an isospin
coupled basis, the $\bar {K} N$ channel can have isospin $I=0$ or
$I=1$, so the resultant $G$-matrices are classified according to
the value of isospin. For $I=0$, $\bar {K} N$ and $\pi \Sigma$ are
the only channels available, while for $I=1$ the $\pi \Lambda$
channel is also allowed. In a schematic notation, each $G$-matrix
fulfills the coupled channel equation:

\begin{eqnarray}
\langle M_1 B_1 \mid G(\Omega) \mid M_2 B_2 \rangle &&=
\langle M_1 B_1
\mid V({\sqrt s}) \mid M_2 B_2 \rangle   \nonumber \\
&& \hspace*{-4cm}+\sum_{M_3 B_3} \langle M_1 B_1 \mid V({\sqrt
s }) \mid
M_3 B_3 \rangle
\frac {Q_{M_3 B_3}}{\Omega-E_{M_3} -E_{B_3}+i\eta} \langle M_3
B_3 \mid
G(\Omega)
\mid M_2 B_2 \rangle \ ,
   \label{eq:gmat1}
\end{eqnarray}
where $\Omega$ is the 'starting energy', given in the lab. frame, and
$\sqrt{s}$ is the invariant center-of-mass energy. In
Eq.~(\ref{eq:gmat1}), $M_i$ and $B_i$ represent, respectively, the
possible mesons ($\bar {K}$, $\pi$) and baryons ($N$, $\Lambda$,
$\Sigma$), and their corresponding quantum numbers, such as coupled
spin and isospin, and linear momentum. The function $Q_{M_3 B_3}$
stands for the Pauli operator preventing the nucleons in the
intermediate states from occupying already filled states.

The prescription for the single-particle energies of all the mesons
and baryons participating in the reaction and in the intermediate
states is written - in nonrelativistic approximation but keeping
relativistic kinematics - as
\begin{equation}
 E_{M_i(B_i)}(p)=\sqrt{p^2 +m_{M_i(B_i)}^2} + U_{M_i(B_i)}
(p,E_{M_i(B_i)}^{qp}(p)) \ ,
\label{eq:spen}
\end{equation}
where $U_{M_i(B_i)}$ is the single-particle potential of each meson
(baryon) calculated at the real quasi-particle energy
$E_{M_i(B_i)}^{qp}$. For baryons, this quasi-particle energy is given
by
\begin{equation}
E_{B_i}^{qp}(p)=\sqrt{p^2 +m_{B_i}^2} +  U_{B_i} (p) \ ,
\label{eq:qpb}
\end{equation}
while, for mesons, it is obtained by solving the implicit equation
\begin{equation}
(E_{M_i}^{qp}(p))^{2}=p^2 +m_{M_i}^2 + {\rm {Re}\,} \Sigma^{ret}_{M_i}
(p,E_{M_i}^{qp}(p)) \ ,\label{eq:qpm}
\end{equation}
where $\Sigma^{ret}_{M_i}$ is the retarded meson self energy.

The $\bar K$ single-particle potential in the Brueckner-Hartree-Fock
approach (at temperature $T=0$) is given by
\begin{equation}
 U_{\bar K}(p_{\bar{K}},E_{\bar K}^{qp})= \sum_{N \leq F} \langle \bar K
N \mid
 G_{\bar K N\rightarrow
\bar K N} (\Omega = E^{qp}_N+E^{qp}_{\bar K}) \mid \bar K N
\rangle \ ,
\label{eq:self0}
\end{equation}
where the summation over nucleon states is limited to the occupied
nucleon Fermi sphere. The ${\bar K}$ self energy is obtained from
the optical potential (neglecting energy derivatives) through the
relation
\begin{equation}
\Sigma^{ret}_{\bar{K}}(p_{\bar{K}},\omega)=2 \
\sqrt{p_{\bar{K}}^2+m_{\bar{K}}^2} \
U_{\bar{K}}(p_{\bar{K}},\omega) \ ,
\label{eq:pik}
\end{equation}
with $\omega=E_{\bar K}^{qp}(p_{\bar{K}})$.  As it can be easily seen
from Eq.~(\ref{eq:self0}), since the $\bar{K}N$ effective interaction
($G$-matrix) depends on the $\bar{K}$ single-particle energy, which in
turn depends on the ${\bar K}$ potential through
Eqs.~(\ref{eq:qpm}),(\ref{eq:pik}), one needs to solve a
self-consistent problem.

After self-consistency is reached, the complete energy-and-momentum
dependent self energy of the ${\bar K}$ can be obtained from
Eq.~(\ref{eq:pik}), which allows one to derive the $\bar K$ propagator
\begin{equation}
D_{\bar{K}}(p_{\bar{K}},\omega)=
\frac{1}{\omega^2-p_{\bar{K}}^2-m_{\bar{K}}^2-
\Sigma^{ret}_{\bar{K}}(p_{\bar{K}},\omega)} \ , \label{eq:prop}
\end{equation}
and the corresponding spectral density, defined as
\begin{equation}
S_{\bar K}(p_{\bar K},\omega) = - \frac {1}{\pi} {\rm Im\,} D_{\bar
K}(p_{\bar K},\omega) \ .
\label{eq:spec}
\end{equation}
We note that our self-consistent procedure amounts to replace the
complex energy dependent self energy,
$\Sigma^{ret}_{\bar{K}}(p_{\bar{K}},\omega)$, in the ${\bar K}$
propagator by that evaluated at the quasi-particle energy,
$\Sigma^{ret}_{\bar{K}}(p_{\bar{K}},\omega=E^{qp}_{\bar K}(p_{\bar
  K}))$.  We denote this procedure as the {\em quasi-particle
  self-consistent approach}, which retains the position and the width
of the peak of the ${\bar K}$ spectral function at each iteration.
This simplification relative to using the complete energy dependence,
as done in Refs.~\cite{Lutz98,Oset00}, allows one to perform
analytically the energy integral of the intermediate loops, thus
reducing the four-dimensional integral equation to a three-dimensional
one.

An important modification for the antikaon properties comes from
considering the pion self energy,
$\Sigma^{ret}_{\pi}(p_{\pi},\omega)$, in the intermediate
$\pi\Sigma$, $\pi\Lambda$ states present in the construction of
the $\bar{K}N$ effective interaction. Since the dressing of the
pions has a strong influence on the antikaon in-medium properties
(cf. Refs. \cite{Oset00,lauran} and also Sections 3.2 and 5 of the
present work), it is of relevance to be more specific on the
actual realization. The pion self energy of the present work
is built up from $ph$, $\Delta h$ and $2p2h$ excitations and
contains the effect of nucleon-nucleon short-range correlations
via a phenomenological model developed in Refs. \cite{OW79,OTW82},
which gives rise to a smooth momentum-dependent Landau-Migdal
parameter $g^\prime$ of the order of 0.6. This is very similar to
the value obtained from microscopic calculations based on the
nucleon-nucleon G-matrix \cite{Dickh1,Dickh2,Dickh3} including the
crossed-channel contributions, similarly as the induced
interaction of Babu and Brown \cite{BB73}. Moreover, the
phenomenological pion self energy used here has been tested in
various reactions involving the interaction of virtual or real
pions with nuclei in Refs. \cite{COF90,COS92,GOS91,CO93}.
Certainly, this pion self energy is extrapolated to higher
densities in the present work and it might not be justified to use
the same value for $g^\prime$. However, the microscopic
study of Ref. \cite{Dickh1}, performed in the context of the
problem of pion condensation at high baryon density, only  showed
a weak dependence of the nucleon-nucleon residual interaction with
density. Nevertheless, we regard the issue of pion dressing at
higher baryon densities still to be open and will present our
G-matrix results in two limits, i.e. with and without pion
dressing.

In order to address energetic nucleus-nucleus collisions, the coupled
$G$-matrix equations in principle have to be extended to
non-equilibrium phase-space configurations. This is quite a formidable
task such that we restrict to finite temperature calculations since
the antikaon dynamics of interest proceeds in the environment of a hot
hadronic fireball.

The introduction of temperature in the $G$-matrix equations affects
the Pauli blocking of the intermediate nucleon states as well as the
dressing of mesons and baryons. The $G$-matrix equation at finite $T$
reads formally as in Eq.~(\ref{eq:gmat1}), but replacing (cf. Ref.
\cite{lauran})
\begin{eqnarray}
Q_{M B} &\rightarrow& Q_{M B}(T) \nonumber \\
G(\Omega) &\rightarrow& G(\Omega,T) \nonumber \\
E_M \ , E_B &\rightarrow& E_M(T) \ , E_B(T) \nonumber \ .
\end{eqnarray}
The function $Q_{M B}(T)$ is unity for meson-hyperon states while, for
$\bar{K}N$ states, it follows $Q_{M B}(T) = 1-n(p_N,T, \mu)$ with the
nucleon occupation number given by
\begin{eqnarray}
n(p_N,T,\mu)=\left[ 1+\exp \left(\frac{E_N(p_N,T)-\mu}{T}\right)
\right ] ^{-1} \ . \label{eq:fermidistribution}
\end{eqnarray}
The chemical potential $\mu$ in Eq.~(\ref{eq:fermidistribution}) is
obtained by imposing the normalization condition for the nucleon
density
\begin{eqnarray}
\rho=\frac{g}{(2\pi)^3} \int  d^3p_N  \ n(p_N,T,\mu) \ ,
\label{eq:density}
\end{eqnarray}
where $g=4$ is the degeneracy factor for symmetric nuclear matter.

A finite temperature also affects the properties of the particles
involved in the process.  The $\bar K$ optical potential at a given
temperature $T$ is calculated according to
\begin{equation}
 U_{\bar K}(p_{\bar{K}},E_{\bar K}^{qp},T)= \int d^3p_N \ n(p_N,T) \ \langle \bar K N \mid
 G_{\bar K N\rightarrow
\bar K N} (\Omega = E^{qp}_N+E^{qp}_{\bar K},T) \mid \bar K N \rangle \
,
\label{eq:self}
\end{equation}
which again is a self-consistent problem for $U_{\bar{K}}$.  More
explicitly, using the partial wave components of the $G$-matrix, we
obtain
\begin{eqnarray}
 U_{\bar{K}}(p_{\bar{K}},E_{\bar K}^{qp},T)&&=
 \frac{1}{2}
\sum_{L, J, I}(2J+1)(2I+1) \int n(p_N,T)\  p_N^2 \ dp_N \\ \nonumber
&& \times  \langle (\bar{K}N) ; \overline{p}| G^{L J I}
(\overline{P},E^{qp}_{\bar{K}}(p_{\bar{K}})+
E^{qp}_{N}(p_{N}),T)
|
(\bar{K} N); \overline{k} \rangle  \ ,
\label{eq:upot1}
\end{eqnarray}
where $\overline{p}$ and $\overline{P}$ are the relative and
center-of-mass momentum, respectively, averaged over the angle
between the external $\bar{K}$ momentum in the lab system,
$p_{\bar{K}}$, and the internal momentum of the nucleon, $p_N$. In
the actual calculations, we include partial waves up to $L=4$. In
extension of the work in Ref.~\cite{lauran}, we here additionally
include the $\Sigma^*(1385)$ resonance dynamics in a similar way
as in Ref.~\cite{jido}. We note that the role of this subthreshold
$\Sigma^*(1385)$ resonance is negligible for ${\bar K}N$ free
scattering observables but, as suggested in Ref.~\cite{Lutz02}, it
can become relevant in the medium since, due to the attraction
felt by the antikaons, one is effectively exploring lower values
of $\sqrt{s}$.

\subsection{Antikaon quasi-particle properties}

With respect to nucleus-nucleus collisions at SIS energies (1--2
A$\cdot$GeV), we have performed the $G$-matrix calculations at a fixed
temperature $T$ = 70 MeV which corresponds to an average temperature
of the 'fireballs' produced in these collisions.  Moreover, we note
that variations in the temperature from 50 - 100 MeV do not sensibly
affect the quasi-particle properties in the medium \cite{lauran}.

\begin{figure}[ht]
  \phantom{a}\vspace*{0cm} \epsfig{file=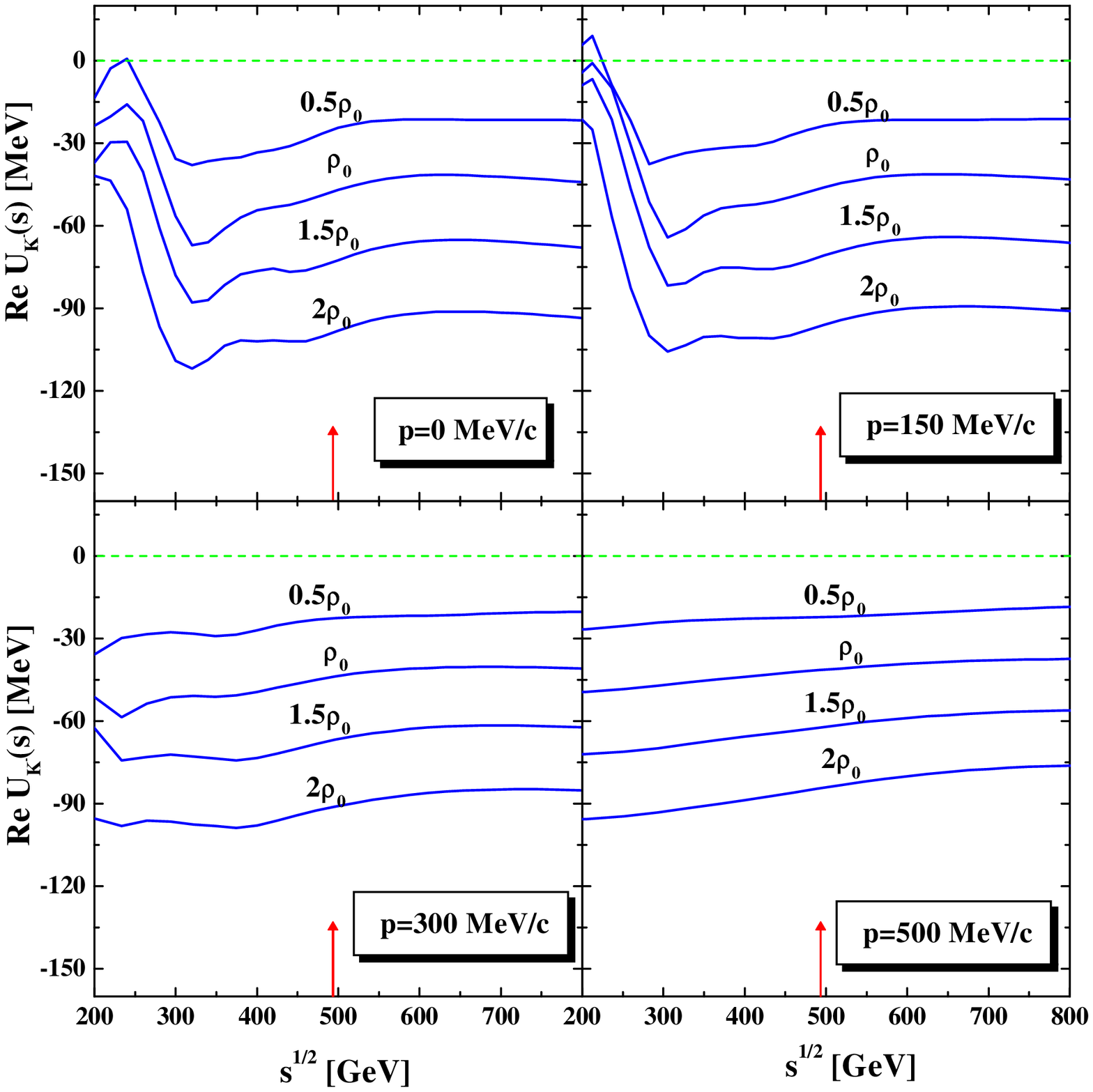,width=15cm}
  \caption{The real part of the antikaon potential Re $U_{\bar K}$ 
   as a function of $\sqrt{s}$ (Eq.~(\protect\ref{sqrt})) for
    different nuclear densities and momenta $p_{\bar K}$ = 0, 150,
    300, and 500 MeV/c, respectively. The arrows show the pole mass of
    the antikaon in free space. }
\label{bild1}
\end{figure}

Whereas in Ref.~\cite{lauran} the antikaon potentials have been shown
at the quasi-particle energy (as a function of momentum and nuclear
density), we display in Fig.~\ref{bild1} the real part of the antikaon
potential Re $U_{\bar K}$ as a function of
\begin{equation}
\label{sqrt}
 \sqrt{s} = \sqrt{\omega^2 - p_{\bar{K}}^2},
\end{equation}
where $\omega$ denotes the antikaon energy and $p_{\bar{K}}$ its
momentum relative to the nuclear matter rest frame, for different
densities and momenta $p_{\bar{K}}$= 0, 150, 300, and 500
MeV/c, respectively. The arrows in Fig.~\ref{bild1} show the pole mass
of the antikaon in free space for orientation. It is seen from
Fig.~\ref{bild1} that the real part of the potential is attractive
throughout and approximately linear in the nuclear density from 0.5
$\rho_0$ to 2 $\rho_0$. Furthermore, the potential is slightly more
attractive for smaller off-shell masses than for masses above the free
invariant mass. With increasing momentum $p_{\bar K}$ the potential
becomes shallower and the dip at an invariant mass of $\sim $ 320 MeV
vanishes. We note, that the results presented in Fig. \ref{bild1} stem
from a $G$-matrix calculation including the pion dressing. When
excluding pion self energies the antikaon potential becomes even more
attractive especially for low invariant masses.

\begin{figure}[ht]
  \phantom{a}\vspace*{0cm} \epsfig{file=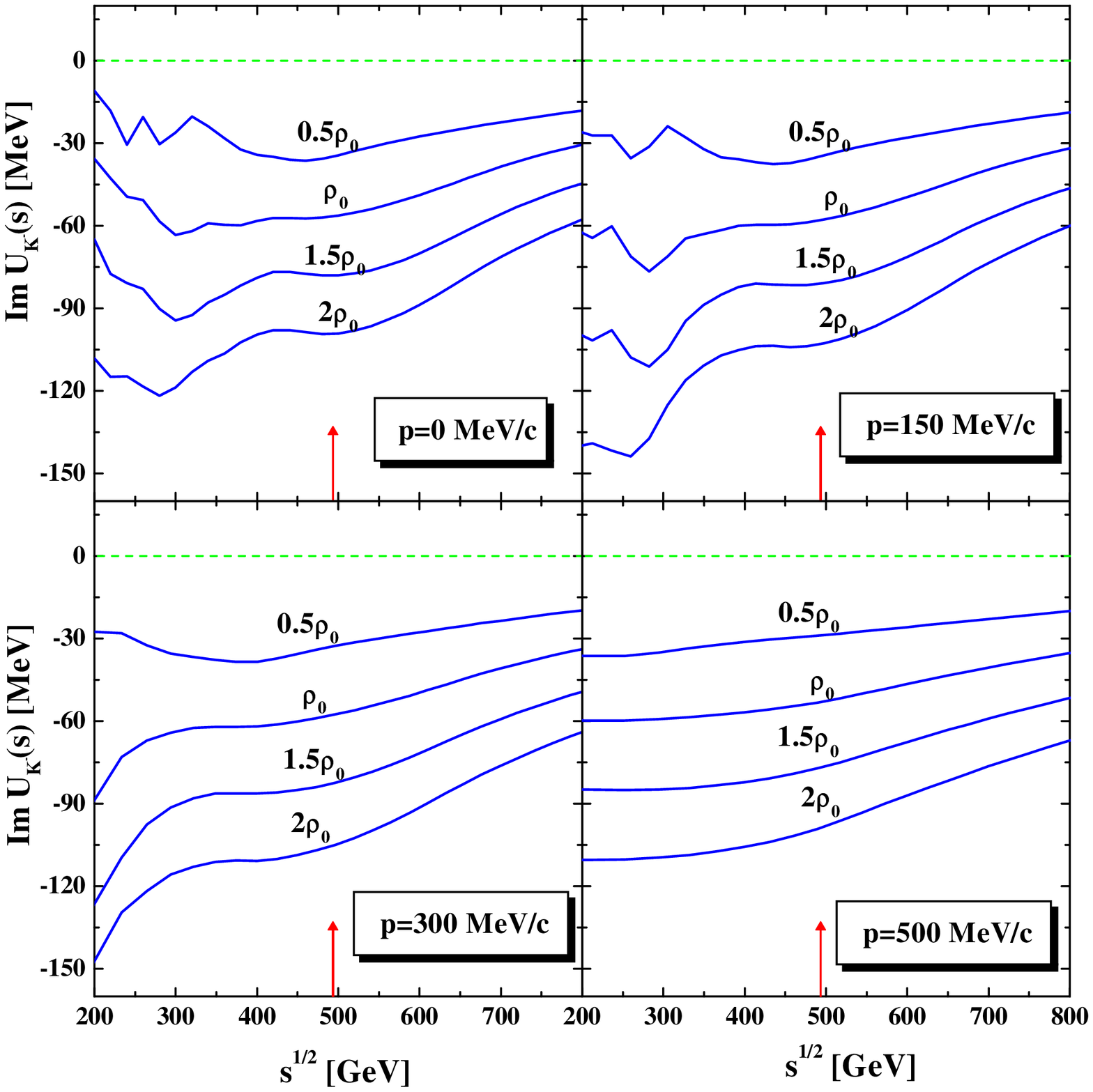,width=15cm}
  \vspace*{0cm} \caption{The imaginary part of the antikaon potential
    Im $U_{\bar K}$ as a function of $\sqrt{s}$ (Eq.~(\protect\ref{sqrt}))
   for different nuclear densities and momenta $p_{\bar K}$ = 0, 150,
    300, and 500 MeV/c, respectively. The arrows show the pole mass of
    the antikaon in free space. } \label{bild2}
\end{figure}

The imaginary part of the antikaon potential Im $U_{\bar K}$ is shown
in Fig.~\ref{bild2} also as a function of $\sqrt{s}$ for different
densities and momenta $p_{\bar K}$ = 0, 150, 300, and 500 MeV/c,
respectively.  The imaginary part is also roughly proportional to the
nuclear density and larger for smaller invariant masses than for
masses higher than the free antikaon mass (arrows). The structure
observed around $\sqrt{s} \sim 300$ MeV at low momentum is due to
resonant $p$-wave coupling to $\Sigma N^{-1}$ states as will be
discussed below.  In general, the imaginary part shows only a weak
dependence on the momentum $p_{\bar{K}}$.

\begin{figure}[ht]
  \phantom{a}\vspace*{0cm} \epsfig{file=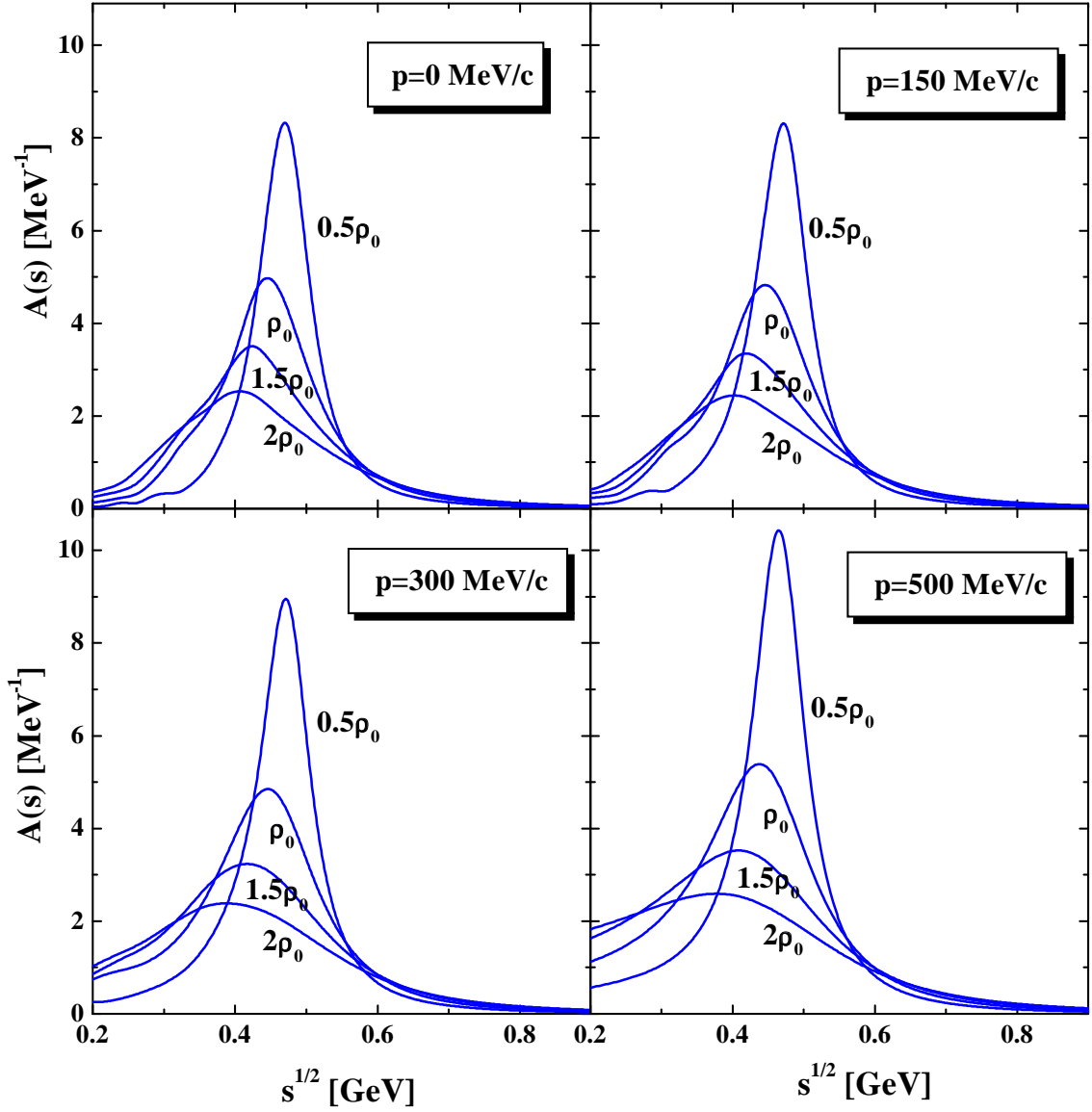,width=15cm}
  \vspace*{0cm} \caption{The antikaon spectral function
    $A(p_{\bar K},\sqrt{s})$ (Eq.~(\protect\ref{spec2})) 
    for different nuclear
    densities and momenta $p_{\bar K}$ = 0, 150, 300, and 500 MeV/c. }
\label{bild3}
\end{figure}

The resulting spectral function $S_{\bar K}(p_{\bar K},\omega)$
is shown in Fig. \ref{bild3} as
\begin{equation}
\label{spec2} A(p_{\bar K},\sqrt{s}) = 2 \
\sqrt{p_{\bar{K}}^2+m_{\bar{K}}^2} \ {S_{\bar K}(p_{\bar
K},\omega)} , \end{equation}
where $\sqrt{s}$ can be identified with the
invariant mass $M$, for different densities and momenta $p_{\bar K}=$
0, 150, 300, and 500 MeV/c, respectively. In line with Figs.~\ref{bild1} and
\ref{bild2}
the maximum of the spectral function shifts to lower invariant
masses with increasing density and becomes substantially broader.
The relative change with momentum $p_{\bar K}$ is only moderate as seen
from Fig.~\ref{bild3}.

As pointed out in Section 2, the information presented in
Figs.~\ref{bild1} and \ref{bild2} -- on a much finer grid in momentum
and density -- is used for the off-shell propagation of antikaons in
the nuclear medium according to the Eqs.~(\ref{eomr}), (\ref{eomp}),
(\ref{eome}) in the transport approach. On the other hand, the
spectral functions from Fig.~\ref{bild3} -- also on a much finer grid
in momentum and density -- enter the collision terms of
Eq.~(\ref{Icoll}) for the production of antikaons in nucleon-baryon,
meson-baryon or pion-hyperon interactions.

\subsection{In-medium transition rates}

Apart from the spectral functions $A_{XP}$ the collision terms in
Eq.~(\ref{Icoll}) depend on the local phase-space densities $N_{X{\vec
    P} M^2}$, that are calculated dynamically for all hadrons, and the
transition matrix-elements squared $|G(({\vec P},M^2) + ({\vec
  P}_2,M_2^2) \rightarrow ({\vec P}_3,M_3^2) + ({\vec
  P}_4,M_4^2))|_{{\cal A,S}}^2$. Here the latter have to be known not
only for on-shell but also for off-shell particles.

For binary reactions involving antikaons like $\bar{K}N \rightarrow
\bar{K}N$ or $\pi \Lambda \leftrightarrow \bar{K}N$ these matrix elements
are determined by the $G$-matrix equation (\ref{eq:gmat1}) such that
no new parameters or unknown cross sections enter the transport
calculations. Actual cross sections are determined as a function of
the invariant energy squared $s$ as
\begin{equation}
\label{cross} \sigma_{1+2 \rightarrow 3+4}(s) = (2 \pi)^5
\frac{E_1 E_2 E_3 E_4}{s} \frac{p'}{p} \ \int d\cos( \theta) \
\frac{1}{(2s_1+1)(2s_2+1)} \sum_i \sum_\alpha \ G^\dagger G,
\end{equation}
where $p$ and $p'$ denote the center-of-mass momentum of the particles
in the initial and final state, respectively. The sums over $i$ and
$\alpha$ indicate the summation over initial and final spins, while
$s_1, s_2$ are the spins of the particles in the entrance channel.
Apart from the kinematical factors, the transition rates are determined
by the angle integrated average transition probabilities
\begin{equation}
\label{crossp} P_{1+2 \rightarrow 3+4}(s) =  \int d\cos(\theta) \
\frac{1}{(2s_1+1)(2s_2+1)} \sum_i \sum_\alpha \ G^\dagger G
\end{equation}
which are uniquely determined by the $G$-matrix elements evaluated for
finite density $\rho$, temperature $T$ and relative momentum $p_{\bar
  K}$ with respect to the nuclear matter rest frame.

\begin{figure}[ht]
  \phantom{a}\vspace*{0cm} \epsfig{file=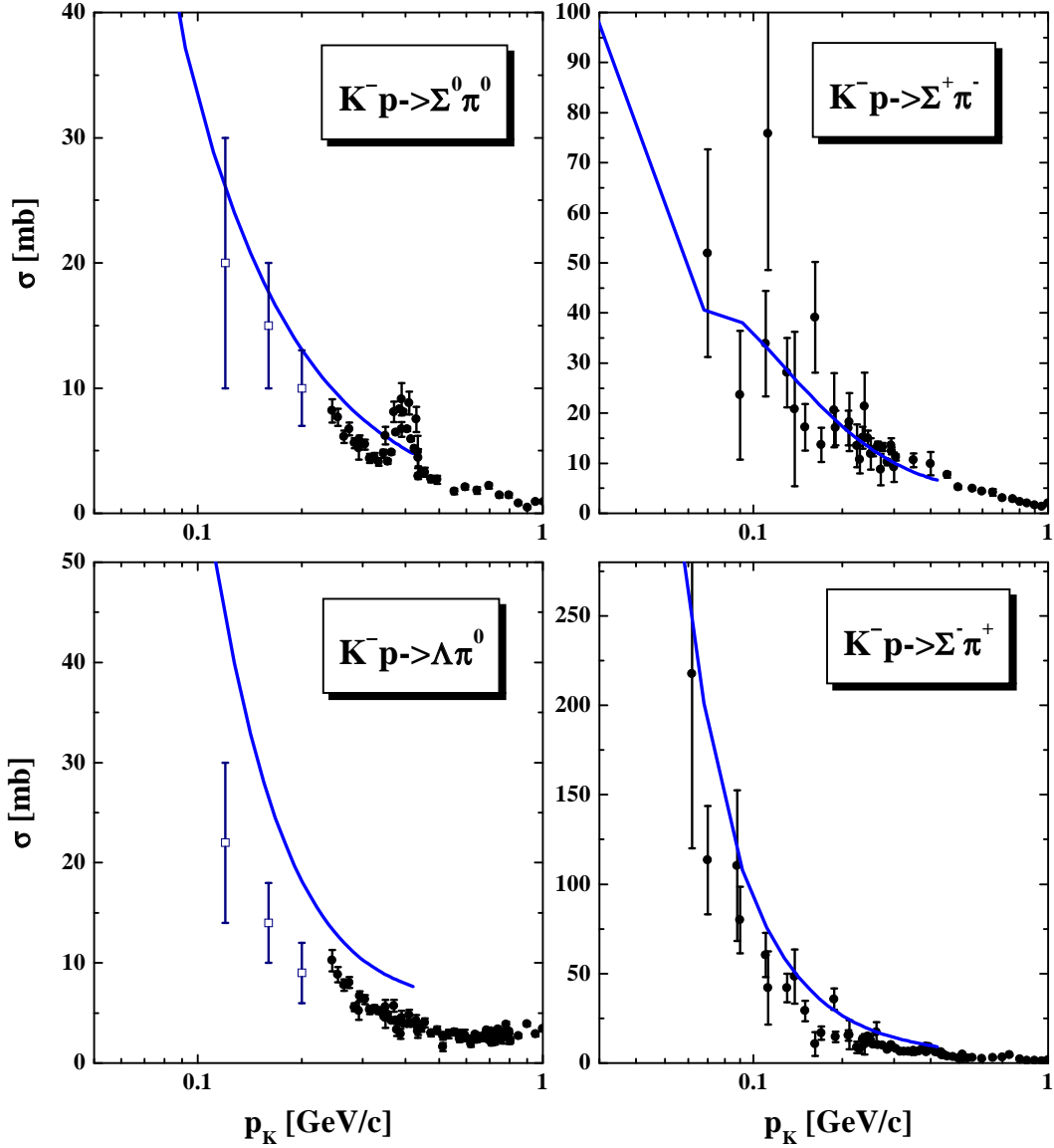,width=15cm}
  \vspace*{-2.5cm} \caption{The $K^- p$ cross sections to the final
    channels $\Sigma^0 \pi^0$, $\Sigma^+ \pi^-$, $\Sigma^- \pi^+$ and
    $\Lambda \pi^0$ according to the $G$-matrix calculations in 'free'
    space in comparison to the data from \protect\cite{LB}.}
\label{bild3b}
\end{figure}

Before coming to the actual in-medium problem we show in Fig.
\ref{bild3b} a comparison of our calculations for the strangeness exchange
cross sections
$K^- p \rightarrow \Sigma^0 \pi^0$, $K^- p \rightarrow \Sigma^+
\pi^-$, $K^- p \rightarrow \Lambda \pi^0$, and $K^- p \rightarrow
\Sigma^- \pi^+$ in 'free space' with the corresponding experimental
data from \cite{LB} as a function of the antikaon momentum $p_{\bar
  K}$ in the laboratory.
As seen from Fig.~\ref{bild3b} we miss the small
resonance for $p_{\bar K} \approx$ 0.4 GeV/c in the $K^- p \rightarrow
\Sigma^0 \pi^0$ channel\footnote{The $\Lambda(1520)$ is not included
  in our calculations.}  and slightly overestimate the $K^- p
\rightarrow \Lambda \pi^0$ channel. Otherwise, the comparison
shows that our transition probabilities are sufficiently realistic
for the vacuum cross sections. It should be pointed out, that the
$\Lambda(1405)$ is generated dynamically in our approach whereas
in Refs. \cite{rr1,rr2} it is treated as an elementary field.

\begin{figure}[ht]
  \phantom{a}\vspace*{0cm} \epsfig{file=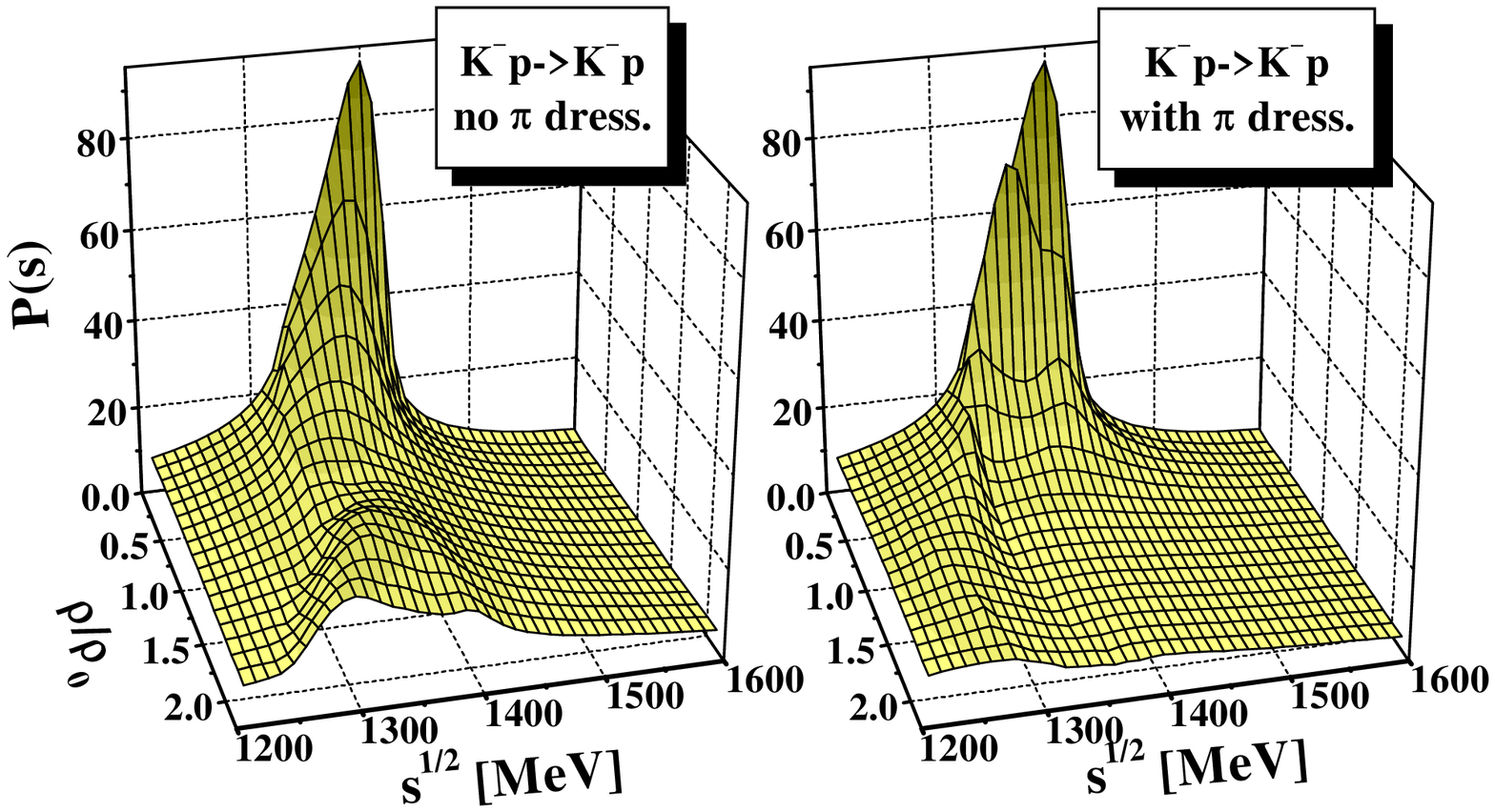,width=15cm}
  \vspace*{1cm} \caption{The transition probability $P_{1+2
      \rightarrow 3+4}(s)$ (Eq.~(\protect\ref{crossp})) for the channel 
   $K^-p \rightarrow K^- p$ as a function of $\sqrt{s}$ and the nuclear
    density $\rho$ (in units of $\rho_0$) for a momentum $p_{\bar
      K}=0$ relative to the nuclear matter rest frame. The l.h.s.
    corresponds to a calculation without pion dressing whereas pion
    dressing is included in the $G$-matrix calculations of the
    r.h.s.} \label{bild4}
\end{figure}

\begin{figure}[ht]
  \phantom{a}\vspace*{0cm} \epsfig{file=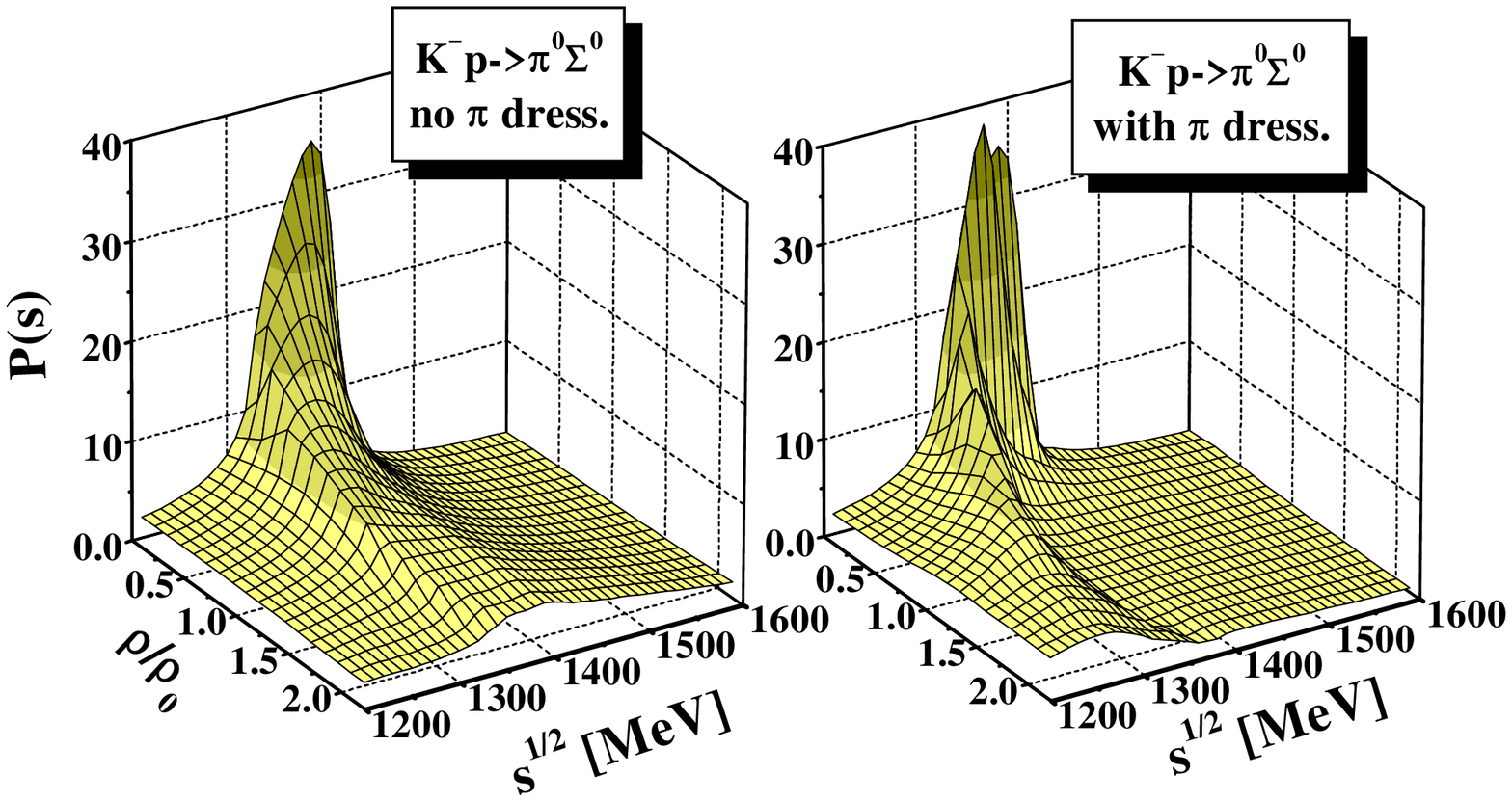,width=15cm}
  \vspace*{1cm} \caption{The transition probability $P_{1+2
      \rightarrow 3+4}(s)$ (Eq.~(\protect\ref{crossp})) for the channel $K^-p
    \rightarrow \Sigma^0 \pi^0$ as a function of $\sqrt{s}$ and the
    nuclear density $\rho$ for a momentum $p_{\bar K}=0$ relative to
    the nuclear matter rest frame. The l.h.s.  corresponds to a
    calculation without pion dressing whereas pion dressing is
    included in the $G$-matrix calculations of the r.h.s.}
  \label{bild5}
\end{figure}

\begin{figure}[ht]
  \phantom{a}\vspace*{0cm} \epsfig{file=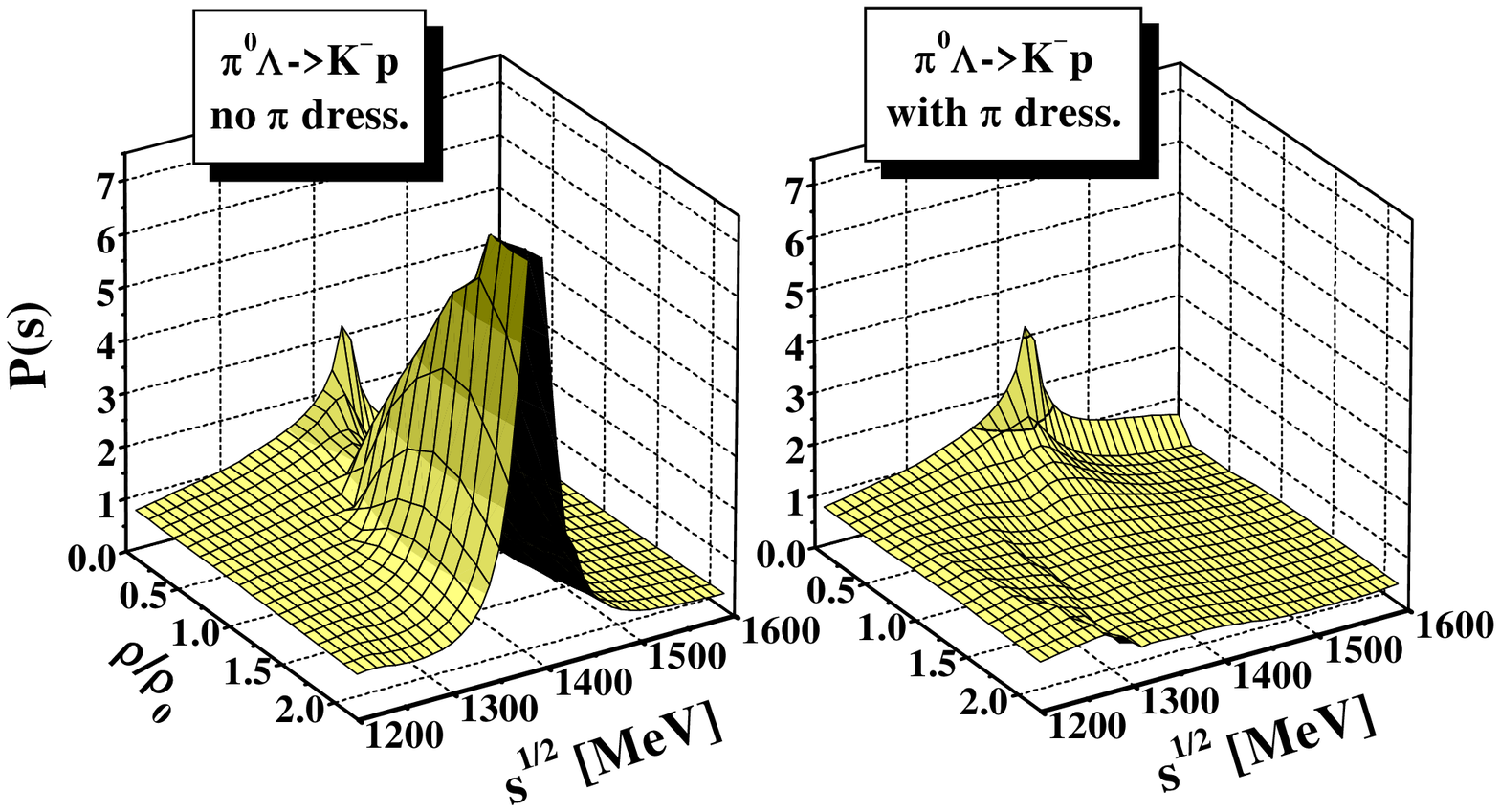,width=15cm}
  \vspace*{1cm} \caption{The transition probability $P_{1+2
      \rightarrow 3+4}(s)$ (Eq.~(\protect\ref{crossp})) for the channel $\Lambda
    \pi^0 \rightarrow K^-p$ as a function of $\sqrt{s}$ and the
    nuclear density $\rho$ for a momentum $p_{\bar K}=0$ relative to
    the nuclear matter rest frame. The l.h.s.  corresponds to a
    calculation without pion dressing whereas pion dressing is
    included in the $G$-matrix calculations of the r.h.s.}
  \label{bild6}
\end{figure}

\begin{figure}[ht]
  \phantom{a}\vspace*{0cm} \epsfig{file=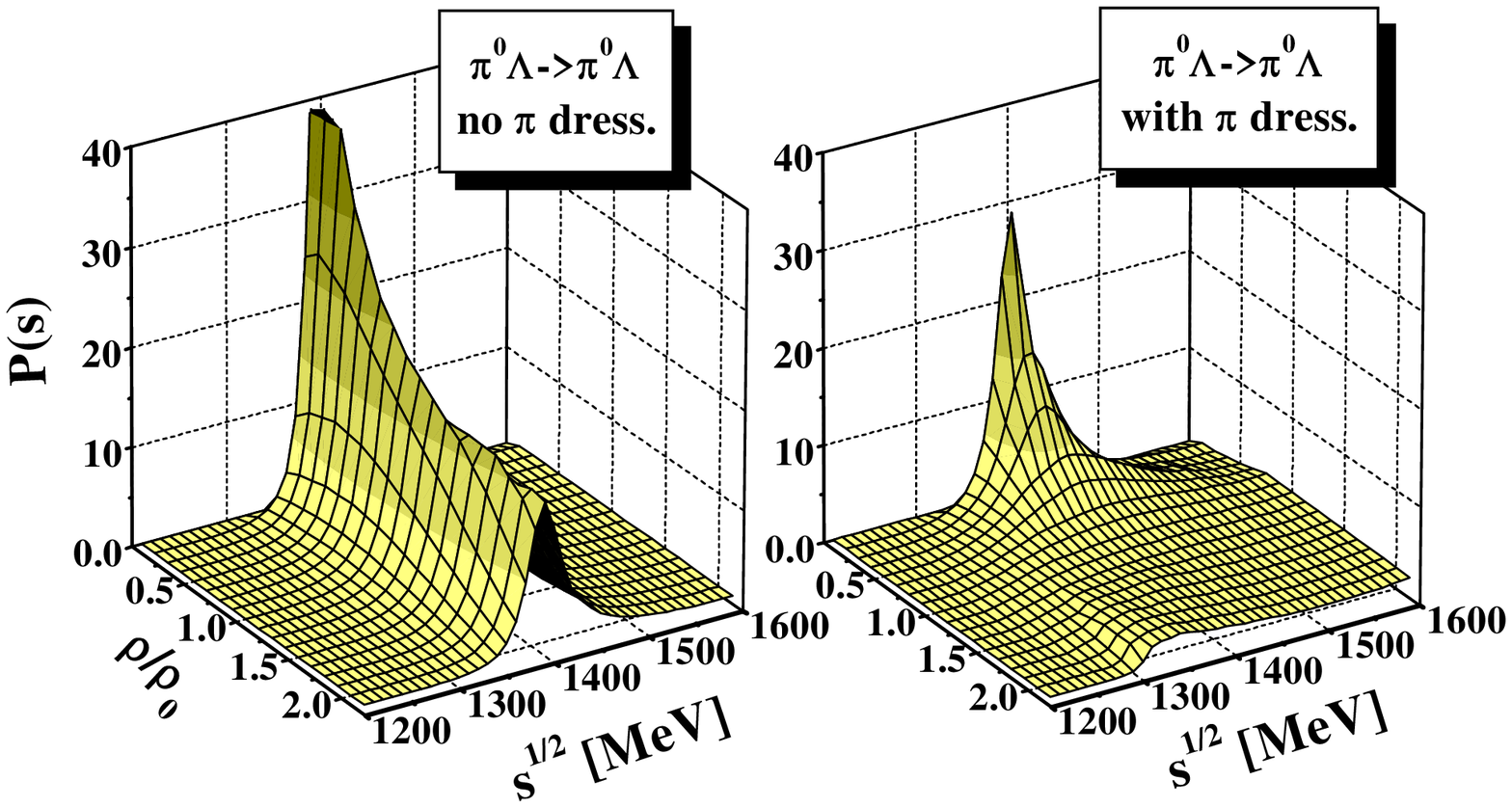,width=15cm}
  \vspace*{1cm} \caption{The transition probability $P_{1+2
      \rightarrow 3+4}(s)$ (Eq.~(\protect\ref{crossp})) for the channel 
 $\Lambda \pi^0 \rightarrow \Lambda \pi^0$ as a function of $\sqrt{s}$ and
    the nuclear density $\rho$ for a momentum $p_{\bar K}=0$ relative
    to the nuclear matter rest frame. The l.h.s.  corresponds to a
    calculation without pion dressing whereas pion dressing is
    included in the $G$-matrix calculations of the r.h.s.}
  \label{bild6b}
\end{figure}

We now turn to the transition probabilities of Eq.~(\ref{crossp}) at
finite density $\rho$, finite temperature $T$ and finite antikaon
momentum $p_{\bar K}$ in the nuclear matter rest frame. They are
displayed in Figs.~\ref{bild4}, \ref{bild5}, \ref{bild6} and
\ref{bild6b} for the reactions $K^- p \rightarrow K^- p$, $K^- p
\rightarrow \Sigma^0 \pi^0$, $K^- p \rightarrow \Lambda \pi^0$ and
$\Lambda \pi^0 \rightarrow \Lambda \pi^0$, respectively. The isometric
plots show the probabilities of Eq.~(\ref{crossp}) as a function of
density $\rho$ (in units of $\rho_0$) and $\sqrt{s}$ for a momentum
$p_{\bar K}=0$ at $T=70$ MeV. The results on the l.h.s. correspond to
a $G$-matrix calculation without pion dressing whereas those on the
r.h.s. include pion dressing as described in Section 3. We note that
the general shape of these transition probabilities does not change
very much at finite momentum $p_{\bar K}$ such that we discard an
explicit representation over this variable.

In order to pick up the physics from
Figs.~\ref{bild4}-\ref{bild6b} we recall that the $K^- p$
threshold in free space corresponds to $\sqrt{s} \approx$ 1.432
GeV, whereas the thresholds for $\pi^0 \Lambda$ and $\pi^0 \Sigma$
are at 1.25 GeV and 1.332 GeV, respectively. In free space ($\rho
= 0$) the coupling to the $\Lambda(1405)$ resonance provides the
dominant matrix elements below (and close to) the ${\bar K}N$
threshold in all channel amplitudes that contain the $I=0$
component (Figs.~\ref{bild4} and \ref{bild5}). Note that, in the
case of off-shell antikaon dynamics in the nuclear medium, also
lower invariant $\sqrt{s}$ become accessible such that a resonant
amplitude slightly below the free threshold will give large
in-medium transition rates.  The same considerations apply to the
$p$-wave $\Sigma^*$ resonance located at 1.385 GeV. Its influence
is seen more clearly in the $I=1$ channel $\Lambda \pi^0
\rightarrow \Lambda \pi^0$ (Fig.~\ref{bild6b}). The reason is
that, since the $p$-wave amplitude is proportional to the product
of center-of-mass momenta in the incoming and outgoing channels, $
p p^ \prime$, it gets enhanced for invariant energies above and
away of the corresponding thresholds. This also explains the
enhanced contribution of this resonance in the channel $K^- p
\rightarrow \Lambda \pi^0$ (Fig.~\ref{bild6}) as density grows due
to the fact that the $K^- p$ threshold, which lies above the
$\Sigma^*(1385)$ in free space, also moves to lower energies in
the medium. This is especially visible on the l.h.s of
Fig.~\ref{bild6} when pion dressing is ignored. In fact, the width
of the $\Sigma^*(1385)$ resonance already increases in the medium
when only the dressing of the antikaons is incorporated due to the
opening of new decay channels, such as $\Sigma^* N \rightarrow
\Lambda N, \Sigma N, \pi \Lambda N, \pi \Sigma N$. However, the
additional dressing of the pions in the $\pi\Lambda$, $\pi\Sigma$
G-matrix intermediate states, to which the $\Sigma^*$ couples very
strongly, enhances tremendously its decay width into the above
mentioned nucleon-induced channels, hence producing a very
'smeared out' contribution, as observed on the r.h.s. of
Fig.~\ref{bild6}.

We also note that in the region of the $\pi\Sigma$ threshold the
squared matrix amplitudes that contain $I=0$ components (see
Figs.~\ref{bild4} and \ref{bild5}) present a double peak structure at
low densities when pion dressing is included.  This was already
pointed out in Ref.~\cite{lauran}, where it was shown that the
in-medium $I=0$ $s$-wave resonance moves down below the $\pi\Sigma$
threshold at normal nuclear matter density $\rho_0$. Hence, as density
increases and, for some range of densities, this resonance shows up as
two distinct bumps.

In any case, it is clear that, whatever resonant structures are
present at low densities, they melt away already at a modest
density of $\sim 0.5 \rho_0$ for all calculations that include
pion dressing (r.h.s.).

Finally, one observes very modest structures, especially visible in
the channels that contain $I=1$ contributions at low energies (see
Figs.~\ref{bild6} and \ref{bild6b}), which explain the behaviour found
in the antikaon potential (cf. Figs. \ref{bild1} and \ref{bild2}) as
well as the bump of the antikaon spectral function (cf. Fig.
\ref{bild3}) around $\sqrt{s} \approx$ 300 MeV, i.e.  roughly 200 MeV
below the free $\bar{K}$ mass. These structures correspond to the
coupling of the ${\bar K}$ meson to $\Sigma$-nucleon hole
configurations.

\section{Antikaon production from $NN$ and $\pi N$ collisions}

Whereas the $G$-matrix approach described in the previous
subsection uniquely determines the off-shell transitions $\bar{K}N
\rightarrow \bar{K} N$ and $\bar{K}N \leftrightarrow \pi Y$, i.e.
the dynamics of hadrons with $s$-quarks in the medium, the
production cross sections of kaons and antikaons from $NN$ or $\pi
N$ collisions are not specified accordingly. As already shown in
Ref. \cite{Cass97} the latter channels are subdominant relative to
the leading $s$-quark exchange reactions $\pi Y \rightarrow
\bar{K} N$. This dominance of the $\pi Y$ reaction channel in
nucleus-nucleus collisions at SIS energies is related to the fact
that kaon cross sections and (due to strangeness conservation) the
hyperon cross sections are about 2 orders of magnitude higher than
the antikaon cross sections \cite{Cass97,Aichelin}. Consequently
$s\bar{s}$-quarks are essentially created with $K Y$ final
channels from $NN$, $\Delta N$ and $\pi N$ reactions and a large
fraction of antikaons stems from the strange flavor exchange
channels $\pi Y \rightarrow \bar{K} N$. We mention that with
increasing bombarding energy also the channel $\pi \pi \rightarrow
K \bar{K}$ may contribute significantly for heavy systems
\cite{Ehehalt,WB}.

For the determination of the in-medium $NN \rightarrow NN K
\bar{K}$ and $\pi N \rightarrow N K \bar{K}$ cross section we here
proceed as follows: from experimental data on the 'free'
production cross sections we extract an average matrix element
squared $|{\cal M}|^2$ by dividing out the phase space for
on-shell particles in the final state and the flux factor (cf.
Ref. \cite{Sibirt}). Using the same matrix element, the off-shell cross
sections then are obtained by employing the dynamical spectral
functions for the hadrons and correcting for the modified final
state phase-space for off-shell hadrons. The actual implementation
is done as follows: In case of meson production by off-shell
baryon-baryon or meson-baryon collisions we either have 2 (e.g.
$\pi N \rightarrow K^+ \Lambda/ \Sigma$), 3 (e.g. $NN \rightarrow
K^+ \Lambda N$ or $K^+ \Sigma N$) or 4 particles (e.g. $NN
\rightarrow NN K^+ K^-$) in the final channel. Since the final
mesons may be off-shell as well, one has to specify the
corresponding mass-differential cross sections that depend on the
entrance channel and especially on the available energy $\sqrt{s}$
in the entrance channel.

\begin{figure}[ht]
  \phantom{a}\vspace*{0cm} \epsfig{file=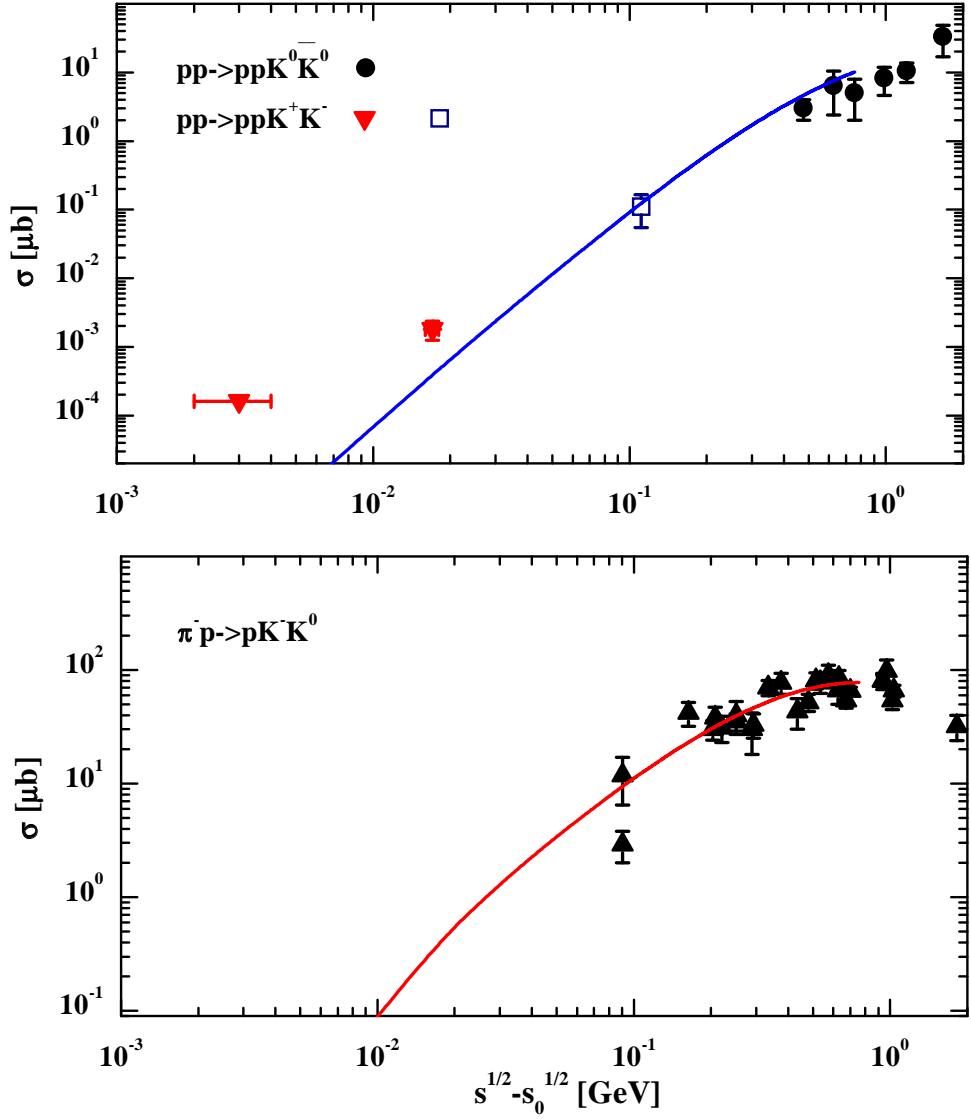,width=15cm}
  \vspace*{-3cm} \caption{The experimental data for the reactions $pp
    \rightarrow K^0\bar{K}^0 pp$, $pp \rightarrow K^+K^- pp$ (from
    Refs. \protect\cite{COSY11,LB,Moscal}) and $\pi^- p \rightarrow
    K^-{K}^0 p$ (from Ref. \protect\cite{LB}) in comparison to our
    parametrizations (solid lines).} \label{bild7}
\end{figure}

We start with the explicit parametrizations for meson ($m$)
production cross sections given in Ref. \cite{CB99} for on-shell
mesons as a function of the invariant energy $\sqrt{s}$ in case of
nucleon-nucleon or pion-nucleon collisions, i.e. $\sigma_{NN
\rightarrow NNm}(\sqrt{s})$ or $\sigma_{\pi N \rightarrow
mN}(\sqrt{s})$, respectively, that are sufficiently well
controlled by experimental data in 'free' space. To this aim we
show in Fig. \ref{bild7} the respective data for $pp \rightarrow
K^0\bar{K}^0 pp$, $pp \rightarrow K^+K^- pp$ from Refs.
\cite{LB,COSY11,Moscal} and $\pi^- p \rightarrow K^-{K}^0 p$ from
Ref. \cite{LB} in comparison to our parametrizations. The reader
should not worry about the fact that the phase-space oriented
parametrization for the $pp$ reactions does not match the 2 lowest
points close to threshold since the latter are enhanced by the
strong final state interaction in the $pp$ and $\bar{K}p$ final
channel in free space. On the other hand, such final state
interactions are essentially 'screened' in the nuclear medium
(especially at high density) such that the phase-space oriented
approximations should hold sufficiently well.

Keeping this concept in mind,  the in-medium  mass
differential cross sections -- far above the corresponding thresholds --
are approximated by
\begin{equation}
\label{c1} \frac{d \sigma_{NN \rightarrow mNN}(\sqrt{s})}{d M^2}
\: = \: \sigma_{NN \rightarrow
mNN}(\sqrt{s^{\phantom{*}}_{\phantom{0}}} -\sqrt{s^{*}_0} \: ) \;
A_m(M^2,P,\rho)
 , \end{equation} where $A_m(M^2, p, \rho)$ denotes
the meson spectral function for given invariant mass $M^2$,
relative momentum $p$ and nuclear density $\rho$ as determined
from $G$-matrix theory in Section 3. In Eq.~(\ref{c1})
the threshold energy $\sqrt{s^*_0} = M_0 + M_1^*+M_2^*$ depends on
the masses of the hadrons in the final channel, i.e. $M_0, M_1^*$
and $M_2^*$. Actual events then are selected by Monte-Carlo
according to Eq.~(\ref{c1}). Close to threshold $\sqrt{s^*_0}$, i.e. for
$\sqrt{s} - M_0 - M^*_1 - M^*_2 \leq 2 \Gamma_{tot}$, where
$M^*_1, M^*_2$  denote the final off-shell masses of two nucleons,
$M_0$ the meson pole mass and $\Gamma_{tot}$ its total width, the
differential production cross section is approximated by a
constant matrix element squared $|{\cal M}_m|^2$ times the available
phase-space,
\be
\label{c2} \frac{d \sigma_{NN \rightarrow mNN}(\sqrt{s})}{d M^2}
\: = \: |{\cal M}_m|^2 \; A_m(M^2,p,\rho ) \;
R_3(s,M^2,M_1^{2*},M^{2*}_2), \ee
where the matrix element $|{\cal M}_m|$ is fitted to the on-shell cross section
typically from 50 to 500 MeV above threshold.
In Eq.~(\ref{c2}) the function $R_3$ denotes the 3-body
phase-space integral \cite{Byc73} in case of a $mNN$ final state.

In case of 4 particles in the final state, i.e. in the channel $NN
\rightarrow K \bar{K} N^*_1 N^*_2$, where the $N^*$'s denote
off-shell nucleons, the differential cross section is approximated
by  \bea &&E_1 E_2 E_3 E_4 \frac{d^{12} \sigma_{BB \rightarrow
NNM_1M_2}(\sqrt{s})}{ d^3p_1 d^3p_2 d^3p_3 d^3p_4} =
\\  \nonumber &&\sigma_{BB \rightarrow NN M_1M_2}(\sqrt{s})
\frac{1}{16 R_4(\sqrt{s})} \delta^4(P_1 + P_2 -p_1-p_2-p_3-p_4),
 \eea
where $R_4$ denotes the 4-body phase-space integral \cite{Byc73}.
Similar strategies have been exploited in case of  subthreshold
$\bar{K}K$ or even $p \bar{p}$ production in proton-nucleus and
nucleus-nucleus collisions in Refs.
\cite{Cass99b,Cass97,Teis1,Sib98}.

\begin{figure}[ht]
  \phantom{a}\vspace*{0cm} \epsfig{file=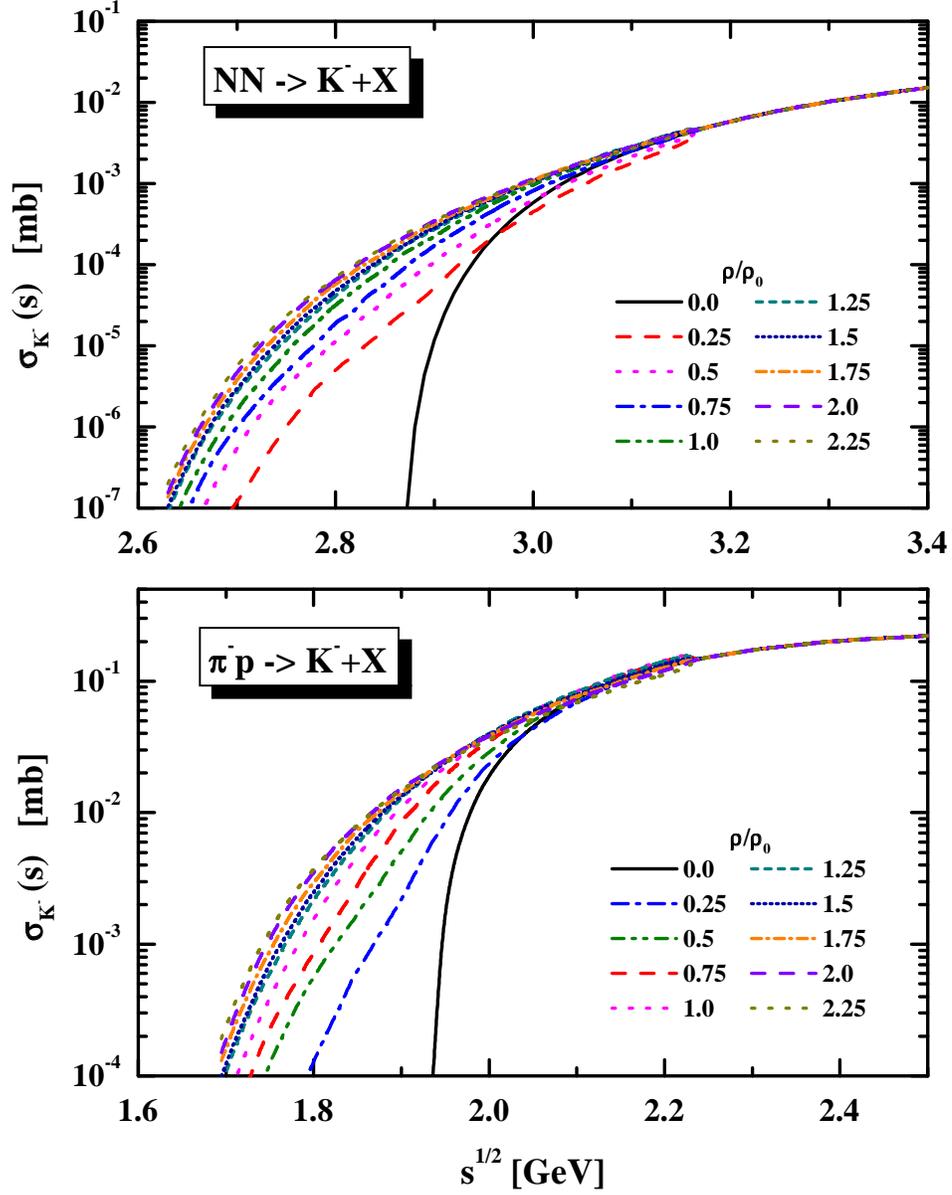,width=15cm}
  \vspace*{-4cm} \caption{The in-medium cross sections for $K^-$
    production from $NN$ and $\pi^- p$ collisions for on-shell
    nucleons and kaons in the final state, however, employing the
    antikaon spectral functions from the $G$-matrix approach in
    Section 3 as a function of $\sqrt{s}$ for different nuclear
    densities ranging from 0.25 $\rho_0$ to 2.25 $\rho_0$ in
    comparison to the cross section in free space (solid lines).}
\label{bild8}
\end{figure}

The resulting cross sections for $K^-$  production from $NN$ and
$\pi^- p$ collisions  using on-shell nucleons and kaons in the
final state, however, employing the antikaon spectral functions
from the $G$-matrix approach in Section 3, are displayed in Fig.
\ref{bild8} as a function of $\sqrt{s}$ for different nuclear
densities ranging from 0.25 $\rho_0$ to 2.25 $\rho_0$ in
comparison to the cross section in free space (solid lines). With
increasing nuclear density the subthreshold production of mesons
becomes enhanced considerably relative to the respective vacuum
cross section due to a shift of the antikaon pole mass and a
broadening of its spectral function (cf. Fig. \ref{bild3}), but
the absolute magnitude stays small below threshold even for $\rho
\approx 2 \rho_0$. We recall again that the $\pi N$ and $NN$
production channels for $K \bar{K}$ pairs are not the leading
channels at SIS energies and minor uncertainties in the off-shell
treatment of the production cross sections are unlikely to show up
in the final $K^-$ abundancies and spectra \cite{Cass97,Aichelin}.

We use isospin symmetry to relate the cross sections for $\pi^- p$
or $pp$ induced reactions to $\pi N$ and $NN$ collisions.
Furthermore, due to a lack of any experimental information, the
$\pi \Delta$ production channel is assumed to be same as the $\pi
N$ channel at the same invariant energy $\sqrt{s}$. The $N \Delta$
and $\Delta \Delta$ production channels again are taken the same
as the $NN$ channel for fixed invariant energy. Apart from
interactions between baryons and pions with baryons also
meson-meson collisions become important with increasing bombarding
energy \cite{Ehehalt,WB}. We include the $\pi \pi \leftrightarrow
K \bar{K}$ channel using the cross section from Ref. \cite{CB99}
and in addition to Ref. \cite{Cass97} also the channel $\pi + K
\leftrightarrow K^*$ employing a Breit-Wigner resonance cross
section with $K^*$ resonance properties from Ref. \cite{PDG}.
Moreover,  the decay $\phi \rightarrow K \bar{K}$ is included in
the transport calculation which, however, is found to be
subdominant.

Now all matrix elements, self energies and/or cross sections for
antikaon production and propagation are specified such that we can
continue with the actual transport calculations. We recall, that
in case of kaons or antikaons  at SIS energies (or charmonia at SPS
energies) we treat the latter
hadrons perturbatively as in Ref. \cite{Cass97,Cabra,brat97}, i.e. each
testparticle achieves a weight $W_i$ defined by the ratio of the
individual production cross section to the total $\pi B$ or $BB$
cross section at the same invariant energy. Their propagation and
interactions are evaluated as for baryons and pions, however, the
baryons (pions) are not changed in their final state when
interacting with a 'perturbative' particle. In case of strange
flavor exchange reactions the individual weight of a $\bar{K}$ is
given to the hyperon $Y$ and viceversa. In this way exact
strangeness conservation can be achieved during the transport
calculation while obtaining reasonable statistics also for
 antikaons at 'subthreshold' energies. We mention again that the actual
(space-time dependent) antikaon width is determined by the
$G$-matrix approach described in Section 3.

\section{Nucleus-nucleus collisions}

\begin{figure}[ht]
  \phantom{a}\vspace*{0cm} \epsfig{file=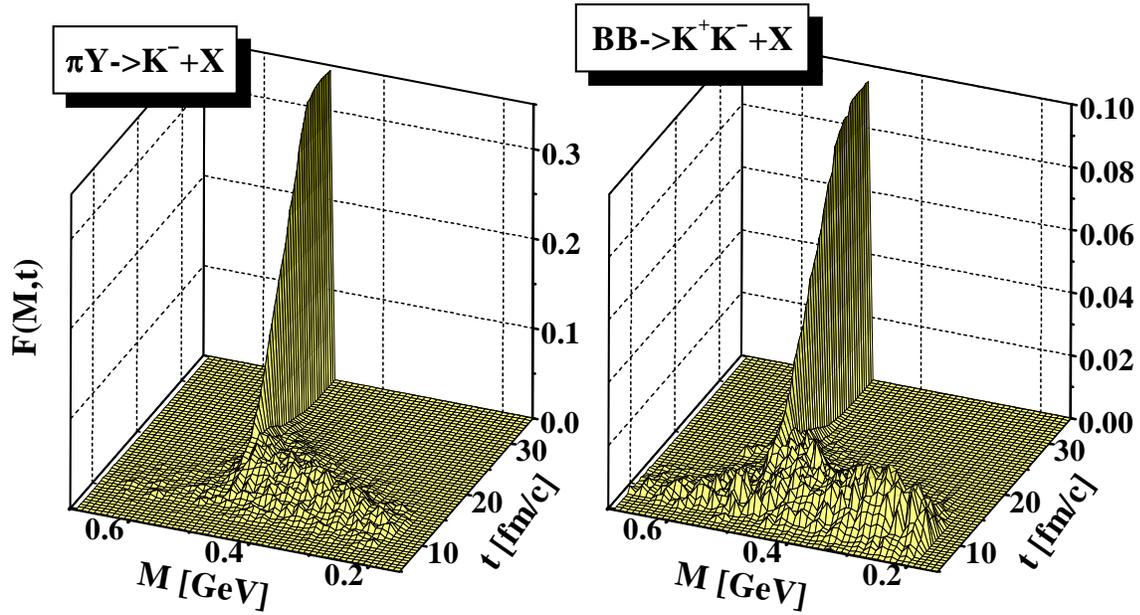,width=15cm}
  \vspace*{1cm} \caption{The time evolution of the antikaon spectral
    distribution (Eq.~(\protect\ref{specshow})) in central $Ni+Ni$
    collisions at 1.8 A$\cdot$GeV. The l.h.s. displays the antikaons
    stemming from the $\pi Y$ channel whereas the r.h.s. shows the
    antikaons emerging from baryon-baryon collisions. } \label{bild9}
\end{figure}

We carry out the concrete applications for nuclear reactions
at SIS energies (1.5 - 2 A GeV)  that have been analysed within
conventional transport models to a large extent (cf. Refs.
\cite{CB99,lix} and Refs. therein). However, before coming to
the actual results we like to address the off-shell propagation of
antikaons and the evolution of the $K^-$ spectral density in time
and invariant mass for central nucleus-nucleus collisions. To this
aim we show in Fig. \ref{bild9} the time evolution of the quantity
\begin{equation}
\label{specshow}
F(M,t) = \int d^3r \int d^3p \ A({\bf
r},t,M^2,{\bf p}) \ N({\bf r},t,M^2,{\bf p})
\end{equation}
as given by the off-shell test-particle approximation
[Eq.~(\ref{testparticle})]. The isometric plots in Fig. \ref{bild9} show
$F(M,t)$ for central collisions of $Ni + Ni$ at 1.8 A$\cdot$GeV
for antikaons stemming from baryon-baryon (BB) (r.h.s.) and $\pi
Y$ reactions (l.h.s.), separately. It is seen from Fig.
\ref{bild9} that the initial distribution of antikaons is widely
spread in invariant mass for both production channels and that --
with the expanding nuclear system -- the antikaons become
practically on-shell for large times, i.e. when moving to 'free'
space with the expanding fireball. In case of the $BB$ production
channel the initial spreading in mass also shows sizeable
contributions for $M \geq$ 0.5 GeV whereas for the $\pi Y$ flavor
exchange reaction dominantly low mass antikaons are produced which
are kinematically not allowed in case of 'free' antikaon masses.
The energy -- to get on-shell finally -- stems from the collective
expansion of the hadronic system.

\subsection{$K^\pm$ spectra at SIS energies}

Since kaons couple only weakly to nucleons and are not absorbed at
low energies their collisional width is rather small such that
they may be treated on-shell to a good approximation as in Ref.
\cite{brat97}. We recall, that the production channel $N\Delta
\rightarrow N K^+ Y$, where $Y$ denotes a hyperon, as well as the
$\Delta \Delta \rightarrow K^+ N Y$ channel is not known
experimentally and simple isospin factors as extracted from pion
exchange \cite{brat97} might not be appropriate. Though there are
some efforts to resolve this uncertainty within extended boson
exchange models \cite{Sib1}, the latter models will hardly be
tested experimentally.  We note that the $N\Delta \rightarrow N
K^+ Y$ and $\Delta \Delta \rightarrow K^+ N Y$ cross sections
employed in the transport models of Refs. \cite{H1,Li1} are larger
by factors of 2 - 3 than ours, which leads to higher $K^+$ (and
hyperon) cross sections from nucleus-nucleus collisions in the
'free' scenario \cite{CB99}. These cross sections are reduced
again in Refs. \cite{H1,Li1} by a repulsive kaon potential in
order to achieve a better agreement with measured kaon cross
sections.

In this work we do not address this particular question in more
detail since only the kaon abundancies are of interest here due to
the associated productions with hyperons $Y = \Lambda, \Sigma$. As
mentioned above, the channel $\pi Y \rightarrow K^- N$ is expected
to be dominant such that the transport calculations have to
reproduce the (experimental) hyperon abundancy with sufficient
accuracy. Since there are no explicit hyperon spectra available
from nucleus-nucleus collisions at SIS energies, we will perform a
detailed comparison to measured $K^+$ spectra (see below).

On the other hand, antikaons couple strongly to nucleons and thus
achieve a large collisional width in the nuclear medium as
demonstrated in Fig. \ref{bild2}. Accordingly,  off-shell
antikaons might be produced at far subthreshold energies (cf. Fig.
\ref{bild8}), become asymptotically on-shell (cf. Fig.
\ref{bild9}) and thus enhance the $K^-$ yield. In fact, as shown
in the model study in Ref. \cite{Cass99b} the $K^-$ yield might be
enhanced up to a factor of 2 for $Ni+Ni$ at 1.8 A$\cdot$GeV when
including the antikaon off-shell propagation. However, in the
latter study the off-shell transition rates $\pi Y \leftrightarrow
K^- N$ had been extrapolated from the on-shell rates in free space
which according to Figs. \ref{bild4}-\ref{bild6b} should be
questionable.

In the following we will show three different limits
simultaneously in comparison to the experimental data: i) a
calculation without any antikaon in-medium effects (denoted by
'free'), where the $G$-matrix elements at density $\rho=0$ are
adopted and the antikaon spectral function is taken as a
$\delta$-function on-mass shell; ii) a calculation using the
$G$-matrix elements, spectral functions and potentials from a
$G$-matrix calculation without pion dressing  and iii) a
calculation using the $G$-matrix elements, spectral functions and
potentials from a $G$-matrix calculation with pion dressing. We
mention, that we use only the $G$-matrix elements calculated at an
average temperature $T$= 70 MeV and do not follow the change of
these matrix elements with decreasing temperature $T$. Apart from
the tremendous numerical effort to calculate the $G$-matrix
additionally on a narrow grid in temperature $T$, the
modifications with temperature are rather moderate or even small
(cf. Ref. \cite{lauran}), such that we discard this variation.

\begin{figure}[ht]
  \phantom{a}\vspace*{-2cm}
  \hspace{1cm} \epsfig{file=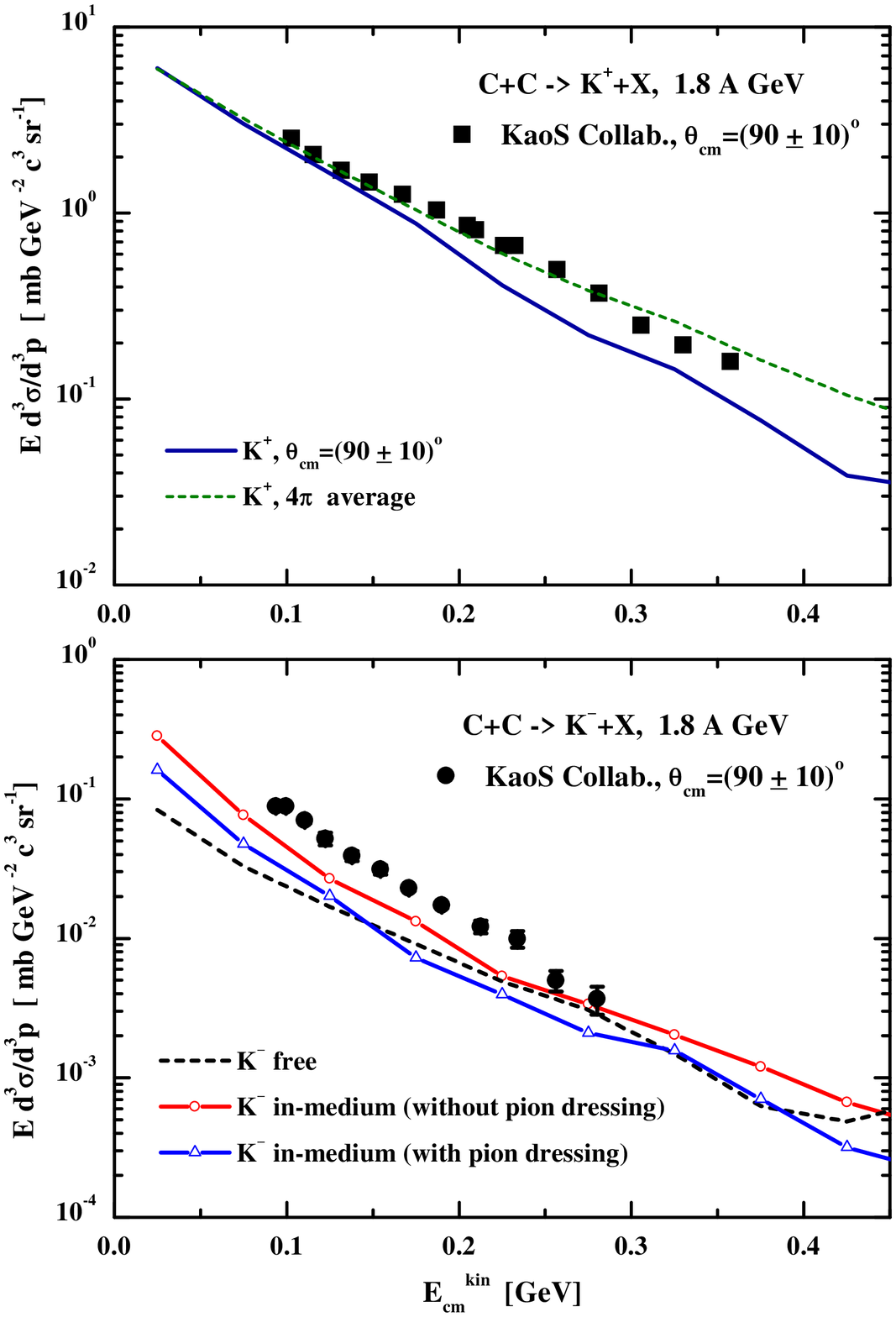,width=12cm} \vspace*{0.0cm}
  \caption{The differential inclusive $K^+$ (upper part) and $K^-$
    spectrum (lower part) for the system $C+C$ at 1.8 A$\cdot$GeV and
    $\theta_{cm} = (90 \pm 10)^o$ in comparison to the data from Ref.
    \protect\cite{K1}. The dashed line in the upper part reflects the
    result of the transport calculation after averaging the $K^+$
    spectra over the angle $\Omega$ in the center-of-mass system,
    while the solid line displays the calculated spectrum for
    $\theta_{cm} = (90\pm 10)^0)$. Lower part: The dashed line
    corresponds to a 'free' calculation, the solid line with open
    triangles to a $G$-matrix calculation including pion dressing
    whereas the solid line with open circles results from a $G$-matrix
    calculation without pion dressing.  } \label{bild10}
\end{figure}

\begin{figure}[ht]
  \phantom{a}\vspace*{-2cm}
  \hspace{0.7cm} \epsfig{file=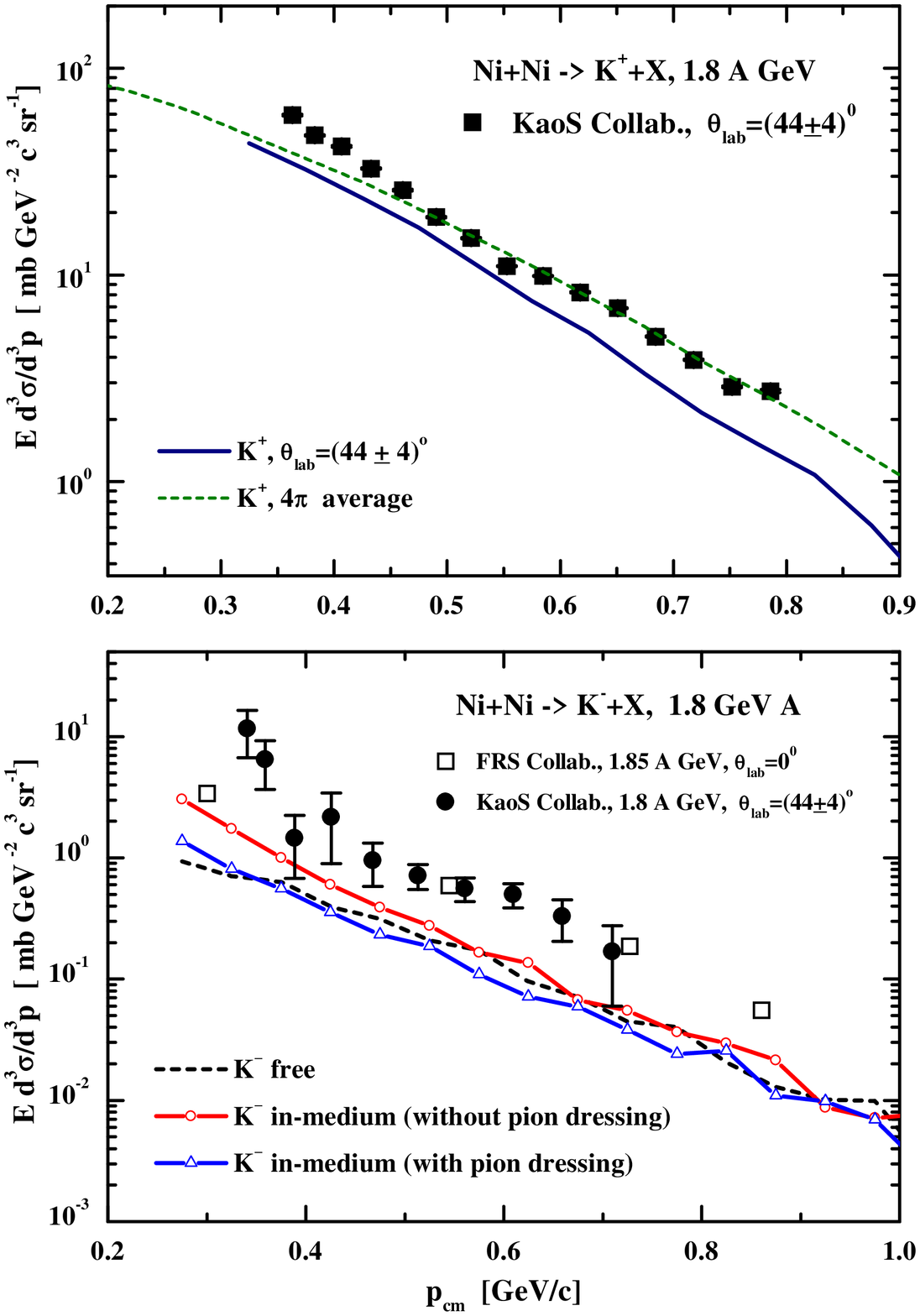,width=13cm} \vspace*{0.0cm}
  \caption{The differential inclusive $K^+$ (upper part) and $K^-$
    spectrum (lower part) for the system $Ni+Ni$ at 1.8 A$\cdot$GeV
    and $\theta_{lab} = (44 \pm 4)^o$ in comparison to the data from
    Refs.  \protect\cite{K2,K3}. The dashed line in the upper part
    reflects the result of the transport calculation after averaging
    the $K^+$ spectra over the angle $\Omega$ in the center-of-mass
    system, while the solid line displays the calculated spectrum for
    $\theta_{lab} = (44\pm 4)^0)$. Lower part: The dashed line
    corresponds to a 'free' calculation, the solid line with open
    triangles to a $G$-matrix calculation including pion dressing
    whereas the solid line with open circles results from a $G$-matrix
    calculation without pion dressing.  } \label{bild11}
\end{figure}

\begin{figure}[ht]
  \phantom{a}\vspace*{-2cm}
  \hspace{1cm} \epsfig{file=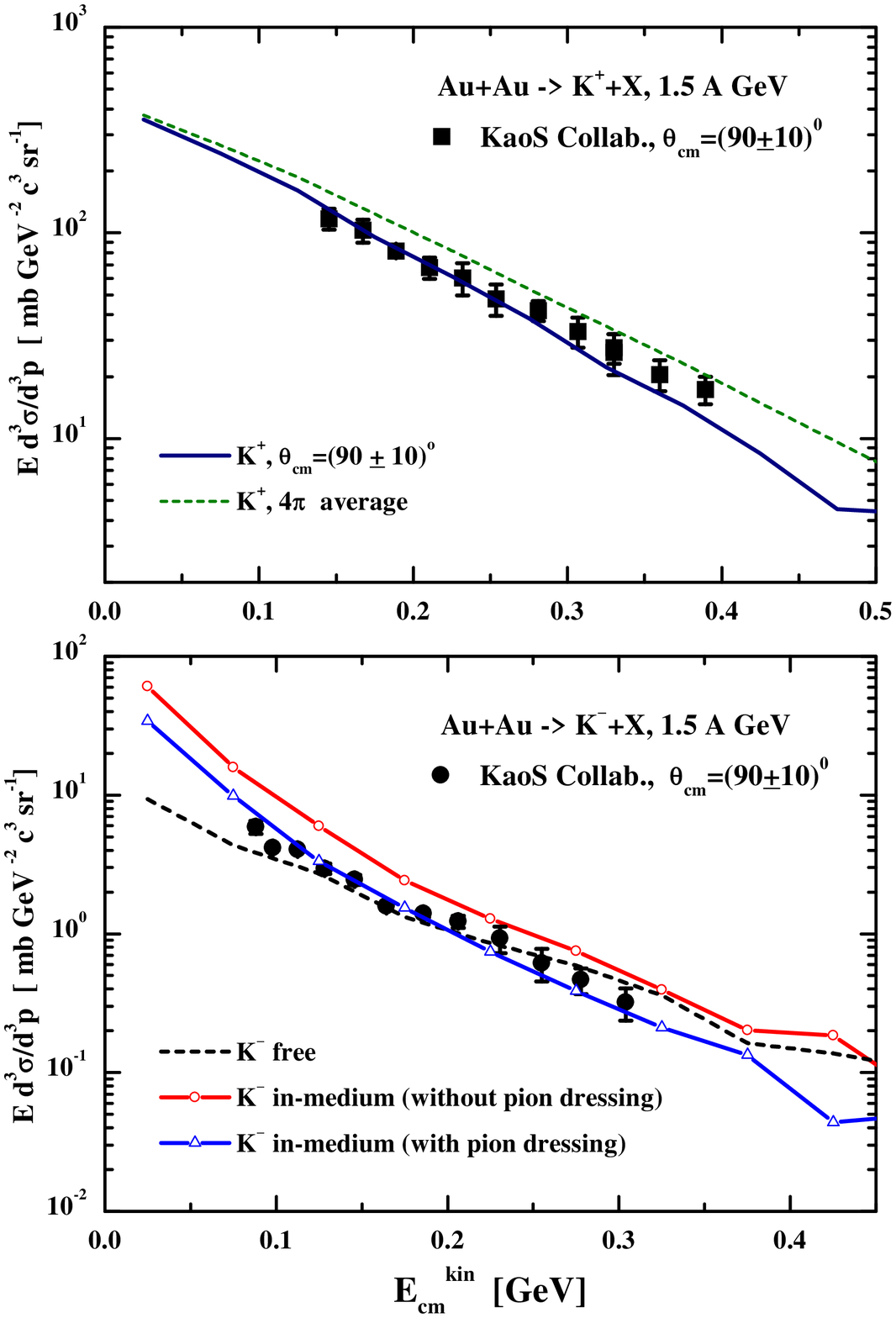,width=12cm} \vspace*{0.0cm}
  \caption{The differential inclusive $K^+$ (upper part) and $K^-$
    spectrum (lower part) for the system $Au+Au$ at 1.5 A$\cdot$GeV
    and $\theta_{cm} = (90 \pm 10)^o$ in comparison to the preliminary
    data from Ref. \protect\cite{K4}. The dashed line in the upper
    part reflects the result of the transport calculation after
    averaging the $K^+$ spectra over the angle $\Omega$ in the
    center-of-mass system, while the solid line displays the
    calculated spectrum for $\theta_{cm} = (90\pm 10)^0)$. Lower part:
    The dashed line corresponds to a 'free' calculation, the solid
    line with open triangles to a $G$-matrix calculation including
    pion dressing whereas the solid line with open circles results
    from a $G$-matrix calculation without pion dressing.  }
  \label{bild12}
\end{figure}

The actual $K^{\pm}$ spectra for the systems $C+C$ at 1.8
A$\cdot$GeV and $\theta_{cm} = (90 \pm 10)^o$, $Ni+Ni$ at 1.8
A$\cdot$GeV and $\theta_{lab} = (44 \pm 4)^o$ and $Au+Au$ at 1.5
A$\cdot$GeV and $\theta_{cm} = (90 \pm 10)^o$ are shown in Figs.
\ref{bild10}-\ref{bild12} in comparison to the data from Refs.
\cite{K1,K2,K3,K4}. The $K^+$ spectra are shown in the upper plots
whereas the $K^-$ spectra are displayed in the lower parts,
respectively. In  Figs. \ref{bild10}-\ref{bild12} the $K^+$
spectra are displayed for the full $G$-matrix calculations only,
since the $K^+$ spectra are practically insensitive to the
different limits addressed for the antikaons within the statistics
of the transport calculations.

In case of $C+C$ (Fig. \ref{bild10}) the 'slope' $T_0$ of the
$K^+$ spectrum, defined by
\begin{equation}
E \frac{d^3 \sigma}{d p^3} \sim \exp(-\frac{E_{cm}}{T_0}),
\end{equation}
is slightly underestimated when comparing the data at $\theta_{cm} = (90\pm10)^o$
with the calculations in the same angular bin (solid line). However, when averaging
the calculated $K^+$
spectrum over the solid angle $\Omega$ in the center-of-mass system (dashed line) the experimental
spectrum is well described in shape as well as absolute magnitude.
This comparison demonstrates that the details of the double
differential cross with respect to the kaon kinetic energy in the cms, $E_{cm}^{kin}$, and angle
$\Omega$ are not fully reproduced by the transport calculation,
but the kaon abundancy and average spectrum compare reasonably well.
Due to the associated production of hyperons (with kaons) we thus
conclude that also the average hyperon abundancy and spectra
should compare sufficiently well with experiment though explicit data are not
available for the $\Lambda$ and $\Sigma$ spectra.

For $Ni+Ni$ at 1.8
A$\cdot$GeV (Fig. \ref{bild11}) the slope of the
$K^+$ spectrum as well as the $K^+$ abundancy is slightly too low
for the calculations at $\theta_{lab} = (44\pm4)^o$ (solid line) in
comparison to the data in the same angular range. Again a very
satisfactory description is obtained in case of the angle averaged
calculated spectrum (dashed line) as in case of the $C+C$ system
in Fig. \ref{bild10}, such that the interpretation and conclusions
drawn in the context of the latter system also hold for $Ni+Ni$ at 1.8
A$\cdot$GeV.
Furthermore, for $Au+Au$ at 1.5 A$\cdot$GeV (Fig. \ref{bild12}) the $K^+$
multiplicity seems to be roughly in line for the spectra in the angular
range $\theta_{cm}= (90\pm 10)^o$ (solid line) as well as for the angular
averaged spectra (dashed line) in comparison to
the preliminary data from Ref. \cite{K4}.

\begin{figure}[ht]
  \phantom{a}\vspace*{-2cm}
  \epsfig{file=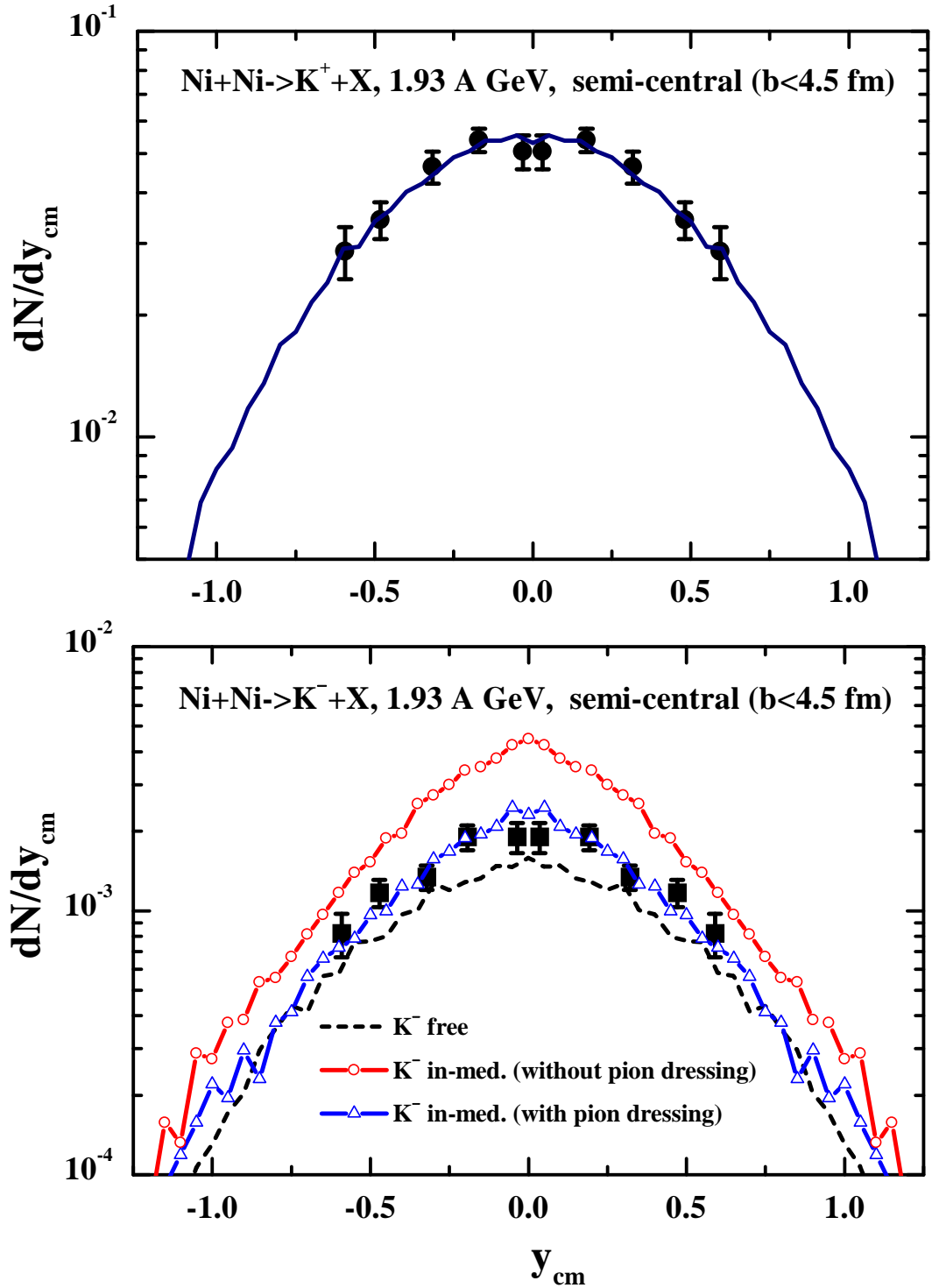,width=14cm} \vspace*{0.0 cm} \caption{The
    rapidity spectrum of $K^+$ (upper part) and $K^-$ mesons (lower
    part) for the system $Ni+Ni$ at 1.93 A$\cdot$GeV and semicentral
    collisions ($b\leq$ 4.5 fm) in comparison to the data from Ref.
    \protect\cite{kaosnew}. The dashed line (lower part) corresponds
    to a 'free' calculation, the solid line with open triangles to a
    $G$-matrix calculation including pion dressing whereas the solid
    line with open circles results from a $G$-matrix calculation
    without pion dressing.  } \label{ni193}
\end{figure}

The KaoS \cite{kaosnew} and FOPI Collaborations \cite{FOPInew}
have independently measured $K^{+}$ production in $Ni+Ni$
collisions at 1.93 A$\cdot$GeV. In fact, their experimental
results for the inclusive $K^{+}$ rapidity distributions agree
quite well in the common region of acceptance in rapidity such
that systematic experimental uncertainties are much better under
control. In Fig. \ref{ni193} we compare our calculated rapidity
distributions for this system with the data from Ref.
\cite{kaosnew} for 'semi-central' collisions that correspond to
impact parameter $b \leq$ 4.5 fm. For $Ni+Ni$ at 1.93 A$\cdot$GeV
our calculations almost perfectly agree with the measured $K^+$
rapidity distribution since deviations in the double differential
$K^+$ spectra with respect to momentum and angle do not show up
any more in the rapidity distributions.

The $K^+$ spectra from our transport calculations thus are found
to be in 'approximate' agreement with the data for the various
systems, however, systematic uncertainties in the order of 30\% in
the $K^+$ (and associated hyperon abundancies) cannot be excluded.
On the other hand, it is presently not clear
if the different data sets are compatible with each other to a
higher accuracy.

We now turn to the results for the $K^-$ spectra (lower parts in
Figs. \ref{bild10}-\ref{ni193}). Whereas for the 'free'
calculations (dashed lines) the experimental $K^-$ spectra are
substantially underestimated for $C+C$ and $Ni+Ni$ at 1.8
A$\cdot$GeV (in line with the calculations in Ref.
\cite{CB99,Cass97}) the preliminary data for $Au+Au$ at 1.5
A$\cdot$GeV in Fig. \ref{bild12} appear to be described rather
well in magnitude. Only the slope of the calculated $K^-$ spectrum
is slightly too hard in this limit. Surprisingly, the $K^-$
rapidity distribution for semi-central collisions of $Ni+Ni$ at
1.93 A$\cdot$GeV (Fig. \ref{ni193}) is underestimated only by less
than a factor of 2 in the 'free' case. This result - on first
sight - appears incompatible with the comparison displayed in Fig.
\ref{bild11} for $Ni+Ni$ at 1.8 A$\cdot$GeV where the $K^-$
spectra are underestimated on average by more than a factor of 4
in the 'free' case\footnote{According to private communication
with the KaoS Collaboration the normalization of the $K^-$ data
for $Ni+Ni$ at 1.8 A$\cdot$GeV is estimated to be too high by up
to a factor of 2, which is on the lower level of the error bars
quoted in Ref. \cite{K2}.}.

The calculated slope $T_0$ of the $K^-$ spectrum  is too high for the
$C+C$ system, roughly compatible for $Ni+Ni$ at 1.8 A$\cdot$GeV
and slightly too high again for $Au+Au$ at 1.5 A$\cdot$GeV in the
'free' scenario when compared to the data in Figs.
\ref{bild10}-\ref{bild12}. The $G$-matrix calculations including
pion dressing (full lines with open triangles) for $C+C$ at
$\theta_{cm} = 90^o \pm 10^o$ are approximately compatible in
magnitude with the 'free' calculations for kinetic energies above 100 MeV,
however, the slope of the
spectrum is now in better agreement with the data. For $Ni+Ni$ at
1.8 A$\cdot$GeV and $\theta_{lab} = 44^o \pm 4^o$ the $G$-matrix
calculations even fall slightly  below the 'free' result whereas
for $Au+Au$ at 1.5 A$\cdot$GeV they are again compatible with the data
and the 'free' result within statistics for $E_{kin}^{cm} >$ 100 MeV.
This also holds for the $K^-$ rapidity distribution for $Ni+Ni$ at
1.93 A$\cdot$GeV in Fig. \ref{ni193} where the full $G$-matrix calculations
 reproduce the experimental spectrum almost 'perfectly'.

In general, the medium modifications of the $K^-$ spectra are
found to be only minor within the $G$-matrix calculations
including pion dressing in the regime of antikaon momenta $p_{cm}$ or
kinetic energies $E_{cm}^{kin}$, where explicit data are available.
This is surprising since the antikaon
spectral functions show a sizeable shift of strength to lower
invariant masses (Fig. 3) such that - according to phase-space -
the cross sections should be enhanced. However, this expectation
is only valid if the transition and production amplitudes do not
change in the medium, which actually does not hold for the present
$G$-matrix calculations as seen from Figs. 5-8. Accordingly, the
dominant production channel $\pi + Y \rightarrow \bar{K} N$
decreases strongly with nuclear density due to the rapid melting
of the $\Lambda(1405)$ and $\Sigma(1385)$ resonances with density
such that less $s$-quarks now can be transferred from hyperons to
antikaons at high baryon density. Qualitatively, our findings are
similar to the self-consistent results of Ref. \cite{Schaffner2}.

The relative role of the hyperon 'resonances' becomes more clear when
looking at the results for the $G$-matrix calculations without
pion dressing (solid lines with open circles in Figs.
\ref{bild10}-\ref{ni193}). In this case all spectra are found to
be enhanced relative to the 'free' calculations and the $G-$matrix
calculations with pion dressing. The experimental spectra for
$C+C$ at $\theta_{cm} = 90^o \pm 10^o$ are still underestimated as
well as for  $Ni+Ni$ at 1.8 A$\cdot$GeV and $\theta_{lab} = 44^o
\pm 4^o$. However, the calculations  for $Au+Au$ at 1.5
A$\cdot$GeV and $Ni+Ni$ at 1.93 A$\cdot$GeV now are clearly above
the measured data. In line with the higher $K^-$ multiplicity also
the spectral slope softens slightly because the antikaon spectral
strength is shifted to lower invariant masses in case of the
'surviving' subthreshold resonances $\Lambda(1405)$ and
$\Sigma(1385)$, respectively. We argue, that the calculations with
and without pion dressing represent a 'band of uncertainty' within
the $G$-matrix calculations, that implicitly depend on the
interaction schemes involved. Nevertheless, in all limits
considered here there is no convincing description of all spectra
simultaneously!

\begin{figure}[ht]
  \phantom{a}\vspace*{-2cm}
  \hspace{1cm} \epsfig{file=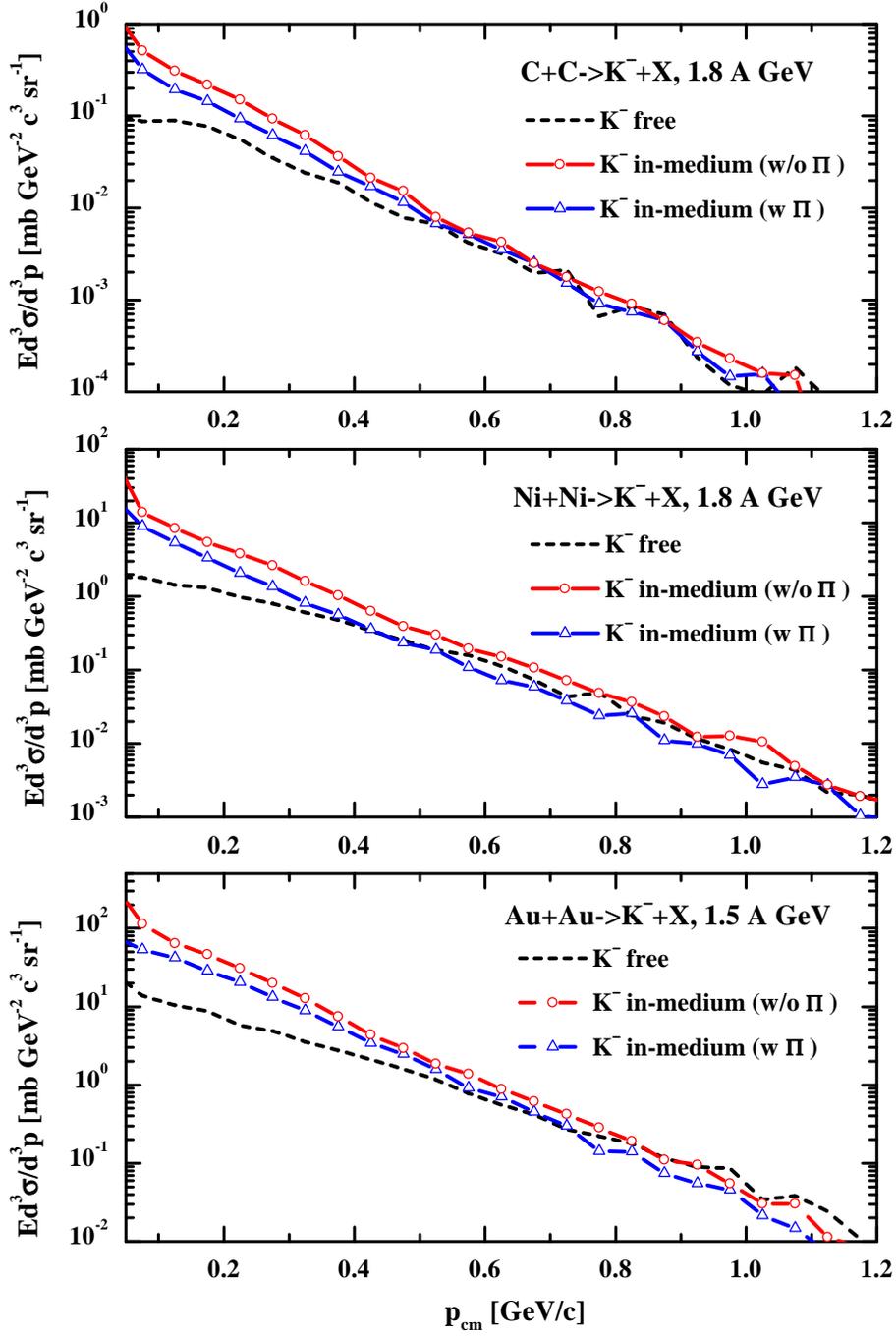,width=12cm} \vspace*{0.0cm}
  \caption{The inclusive momentum spectrum of $K^-$ mesons for the
    systems $C+C$ $Ni+Ni$ and $Au+Au$ as a function of the antikaon
    momentum in the center-of-mass system $p_{cm}$.  The dashed lines
    correspond to a 'free' calculation, the solid lines with open
    triangles to $G$-matrix calculations including pion dressing
    whereas the solid lines with open circles result from $G$-matrix
    calculations without pion dressing.  } \label{pcm}
\end{figure}

The question thus arises if medium effects for antikaons
show up in kinematical regimes not accessible to the present
detector setups. To this aim we show in Fig. \ref{pcm} the Lorentz
invariant $K^-$ spectra for $C+C$ and $Ni+Ni$ at 1.8 A$\cdot$GeV
and $Au+Au$ at 1.5 A$\cdot$GeV as a function of the antikaon
momentum $p_{cm}$ in the cms frame. Whereas above $\sim$ 0.5 GeV/c all
spectra within the limits addressed before are roughly comparable,
a large enhancement at low momenta is found for all systems within
the $G$-matrix calculations relative to the 'free' case. This
enhancement is even more pronounced when excluding pion dressing
(full lines with open circles). The $K^-$ enhancement at low
momenta is essentially due to a deceleration of antikaons in the
combined attractive Coulomb and nuclear mean field as well as due
to the off-shell propagation of 'low mass' antikaons. In the off-shell
propagation (see Eqs. (6)-(8))
spectral components with masses below the free pole mass (cf. Fig.
11) decrease in momentum during the expansion phase according to Eq. (7)
in order to become on-shell in free space.

The results from Fig. \ref{pcm}, which qualitatively agree with the earlier
analysis by Wang et al. \cite{wang98} using on-shell transport calculations,
   also explain to some extent why the calculations
within the full $G$-matrix approach are compatible with the $K^-$
rapidity spectra for $Ni+Ni$ at 1.93 A$\cdot$GeV (Fig.
\ref{ni193}), that experimentally have been obtained by
extra\-polation to low antikaon momenta, but not with the $K^-$
spectra for $Ni+Ni$ at 1.8 A$\cdot$GeV (Fig. \ref{bild11}) in the
higher momentum range of the KaoS acceptance. The reason is most
likely related to the strong increase of the antikaon spectrum at
low momenta in the cms.

\subsection{$K^-/K^+$ ratios versus centrality}
\begin{figure}[ht]
  \phantom{a}\vspace*{0cm}
  \epsfig{file=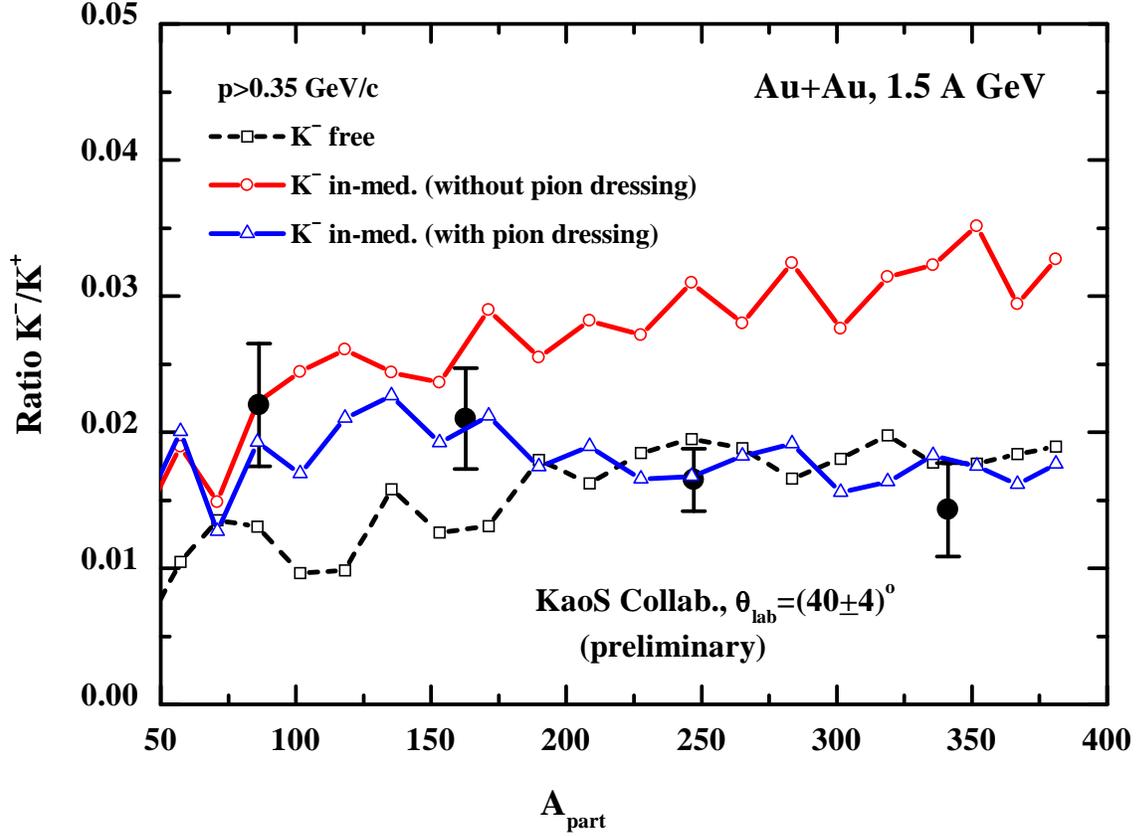,width=15cm} \vspace*{1cm} \caption{The
    $K^-/K^+$ ratio as a function of centrality, expressed in terms of
    the number of participating nucleons $A_{part}$, for the system
    $Au+Au$ at 1.5 A$\cdot$GeV and $\theta_{cm} = (90 \pm 10)^o$
    including a cut in momentum $p_{cm} \geq 0.35$ GeV/c.  The
    preliminary experimental data have been taken from Ref.
    \protect\cite{K4}.  The dashed line corresponds to a 'free'
    calculation, the solid line with open triangles to a $G$-matrix
    calculation including pion dressing whereas the solid line with
    open circles results from a $G$-matrix calculation without pion
    dressing. The fluctuations of the lines are due to the limited
    statistics. }
\label{bild13}
\end{figure}

We continue with the $K^-/K^+$ ratio in $Au+Au$ collisions at 1.5
A$\cdot$GeV as a function of centrality which has been also
addressed experimentally by the KaoS Collaboration \cite{K4}. In
Fig. \ref{bild13} we show this ratio as a function of the number
of participating nucleons $A_{part}$
 for the three
limits discussed before including a cut $p_{cm} > 0.35$ GeV/c which
roughly corresponds to the acceptance of the KaoS spectrometer in Ref. \cite{K4}.
In the 'free' calculations as well as $G$-matrix calculations without pion dressing
this ratio -- within statistics --  increases with
centrality or $A_{part}$. This increase is no longer present
for the $G$-matrix calculations with pion dressing
(full line with open triangles).
The preliminary data for this
ratio from the KaoS Collaboration \cite{K4} (full circles) show - within errorbars -
a slight decrease with $A_{part}$ or  an approximately constant
value of $\sim$ 2\%.  This experimental value is not compatible
with the $G$-matrix calculation excluding pion dressing as well as the
'free' calculation, however, in the order of the full $G$-matrix results.

Some comments on the $K^-/K^+$ ratio appear necessary since the
dependence on centrality and its actual values strongly correlate
with the angle and momentum cut introduced. This comes about
because the $K^-$ spectra show a softer slope in the momentum
spectrum relative to the $K^+$ mesons, which we address to
attractive in-medium effects for the antikaons. An alternative or
complementary interpretation persists in relating the different
slopes to different 'freeze-out' times since $K^+$ mesons decouple
early in view of their low cross section with baryons, whereas
antikaons interact strongly up to rather low densities of the
expanding hadronic fireball. Consequently the relative change of
the $K^-$ to $K^+$ spectral slopes with centrality strongly affect
the $K^-/K^+$ ratio in Fig. \ref{bild13} (for a high momentum cut)
such that no 'simple' interpretation can be drawn directly.
Without explicit representation we mention that in our
calculations the $K^-/\pi^ -$ and $K^+/\pi^+$ ratios increase
both with centrality (or $A_{part}$). However, also the difference
in the slopes for $K^+$ and $K^-$ mesons increases with $A_{part}$
for the full $G$-matrix calculations including pion dressing.

\subsection{$K^\pm$ angular distributions}
We finally come to the $K^{\pm}$ angular distributions in the cms
for $Au+Au$ collisions at 1.5 A$\cdot$GeV that also have been
measured by the KaoS Collaboration \cite{Andreas} for two
different centrality cuts. In Fig. \ref{final} we show these
angular distributions  for the three limits discussed before for
$K^-$ mesons (lower part), whereas the angular distributions for
$K^+$ mesons (upper part) are shown for the full $G$-matrix
calculations, only, since the same distribution has been obtained
in the three limits (within statistics). All angular distributions
are normalized to unity for $\cos \theta_{cm}$ = 0.

We see that the $K^+$ angular distributions from our calculations
for semi-central ($b <$ 6 fm) and non-central ($b
>$ 6 fm) collisions are more isotropic than the data. This points
towards a lower amount of $K^+$ rescattering in the data
\cite{Andreas} especially for non-central collisions. On the other
hand, all three limits are compatible with the experimental
measurements for $K^-$-mesons in semi-central reactions, which are
roughly compatible with an isotropic distribution within
statistics. Differences appear only for non-central reactions,
where the 'free' (dashed line) and $G$-matrix calculation without
pion dressing (solid line with open circles) are closer to
isotropy while the full $G$-matrix calculations with pion dressing
(solid lines with open triangles) show a distribution, that is
forward-backward peaked in the cms in better agreement with the
data. We attribute this result to the lower amount of $s$-quark
exchange reactions in the full $G$-matrix calculations (cf.
Section 3).

\section{Summary}
In this work we have studied the production and propagation of
antikaons with dynamical spectral functions $A_h(X,\vec{P},M^2)$
using off-shell transport theory.  The in-medium properties of the
antikaons have been determined in a coupled-channel $G$-matrix
approach including elastic scattering, charge exchange and
$s$-quark exchange reactions with baryons. Antikaon mean-field
potentials as well as spectral properties are uniquely determined
within the $G$-matrix approach. However, the actual results
strongly depend on the many-body scheme involved, especially on
the dressing of the pions. Though in all cases the spectral
strength of the antikaon is shifted to lower invariant masses with
increasing density, we find that, when including pion
self energies, the $\Lambda(1405)$ and $\Sigma(1385)$ resonance
structures in the transition probabilities melt away with baryon
density already at $\sim$ 0.25 $\rho_0$. This implies that $K^-$
absorption as well as production from pion-hyperon collisions
should be strongly suppressed in nuclear media as produced in
nucleus-nucleus collisions at SIS energies (1--2 A$\cdot$GeV) or
even proton-nucleus reactions at 2--3 GeV \cite{scheinast}.

\begin{figure}[ht]
  \phantom{a}\vspace*{0cm}
  \epsfig{file=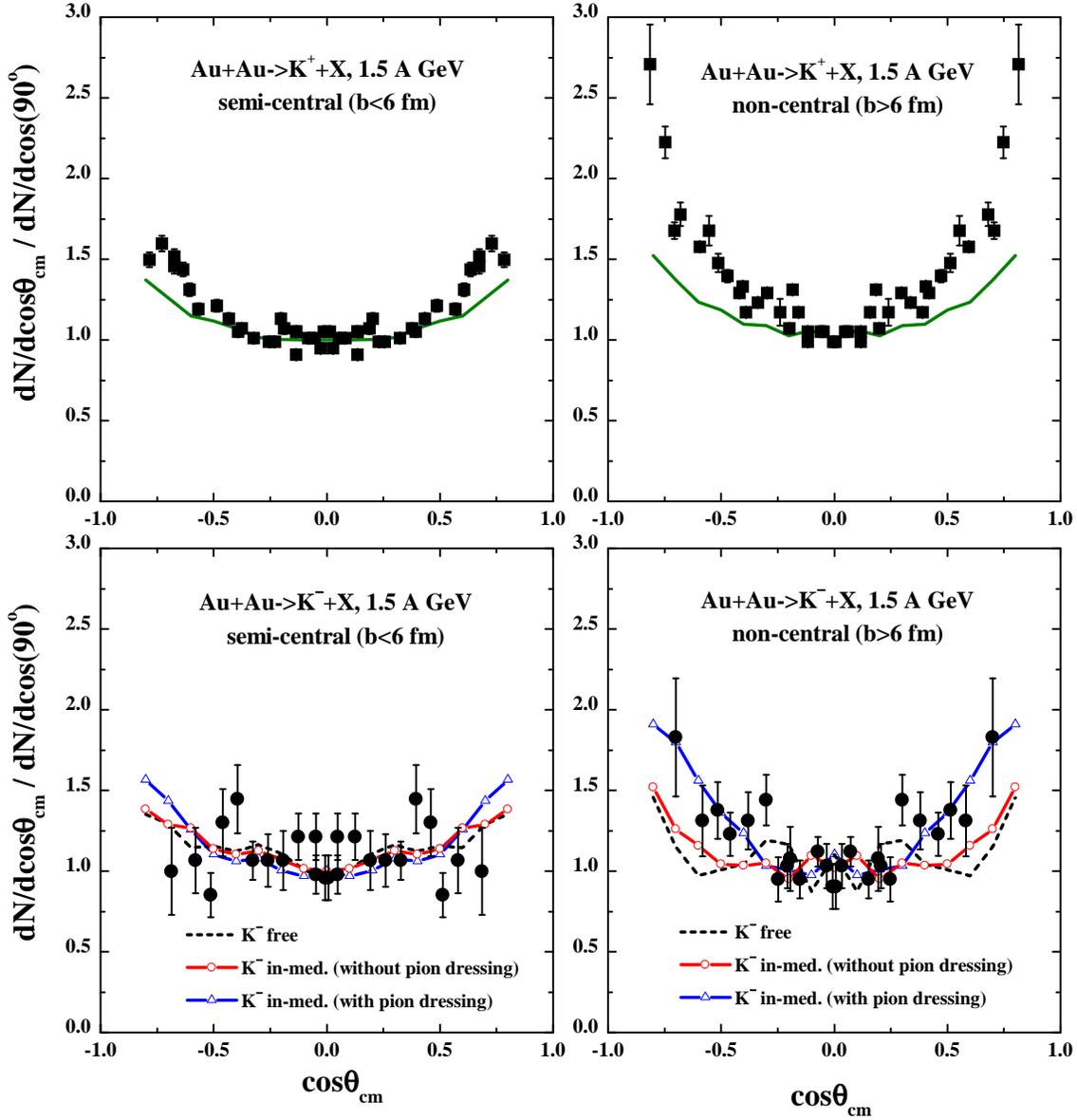,width=15cm} \vspace*{1cm} \caption{The
    $K^+$ (upper part) and $K^-$ angular distribution (lower part)
    for semi-central (l.h.s.) and non-central (r.h.s.)
    $Au+Au$ collisions at 1.5 A$\cdot$GeV. All angular distributions
   are normalized to unity for $\cos \theta_{cm}$ = 0.
  The preliminary experimental data have been taken from Ref.
    \protect\cite{Andreas}.  The dashed line corresponds to a 'free'
    calculation, the solid line with open triangles to a $G$-matrix
    calculation including pion dressing whereas the solid line with
    open circles results from a $G$-matrix calculation without pion
    dressing. The fluctuations of the lines for the $K^-$ angular
    distributions are due to the limited
    statistics again. }
\label{final}
\end{figure}

From our dynamical calculations we find that the experimental
$K^-$ spectra for $^{12}C + ^{12}C$ and $^{58}Ni + ^{58}Ni$ are
underestimated in the 'free' as well as full $G$-matrix transport
approach whereas the preliminary spectra for $Au + Au$ at 1.5
A$\cdot$GeV are rather well described in case of calculations
without any medium effects as well as with all medium effects
included. The dominant medium effect is a 'softening' of the
antikaon spectra which is more in line with the data. Furthermore,
the full in-medium calculations agree very well with the $K^-$
rapidity spectra for semi-central collisions of $Ni+Ni$ at 1.93
A$\cdot$GeV while the spectra are underestimated by up to a factor
of 2 for 'free' transition matrix elements and spectral functions.
In addition, the centrality dependence of the $K^-/K^+$ ratio for
$Au+Au$ reactions at 1.5 A$\cdot$GeV is rather well in line with
the preliminary data of the KaoS collaboration for our full
$G$-matrix calculations including pion dressing. The latter limit
also provides the best description for the $K^-$ angular
distributions in the cms, especially for  non-central $Au+Au$
reactions at 1.5 A$\cdot$GeV (Fig. \ref{final}), which we
interpret as a consequence of the lower amount of $s$-quark
exchange reactions in the full $G$-matrix calculations.

Nevertheless, in all limits considered in this work there is no
convincing description of all spectra simultaneously. This failure
might be either attributed to larger systematic errors in the
experimental data -- as found for $K^+$ production in $p+A$
reactions from 1--2.5 GeV \cite{zibbi} -- or to an inadequacy of
the many-body schemes adopted for the coupled-channel $G$-matrix
calculations, especially the 'pion dressing'.  Additionally, there
is no direct experimental test of the hyperon dynamics
incorporated in the off-shell transport approach since no explicit
hyperon spectra are available for the systems measured at the SIS
accelerator. These issues can only be settled by new experimental
data in a wider kinematical regime with emphasis on low antikaon
momenta in the center-of-mass frame since the in-medium effects
are found to be most pronounced in this region as established
experimentally also in $K^+$ production from $p+A$ reactions for
bombarding energies from 1-2 GeV \cite{Neki}.

\vspace{1cm} The authors like to thank M. Lutz for stimulating
discussions and P. Senger for valuable comments on the manuscript.  L.T.
wishes to acknowledge the hospitality extended to her at the Institut
f\"ur Theoretische Physik (Universit\"at Giessen). This work is also partly
supported by the DGICYT project BFM2002-01868 and by the Generalitat
de Catalunya project SGR 2001-64.


%
%
%
\newpage
%
%

%

\clearpage

\clearpage

\clearpage

\clearpage

\clearpage

\clearpage

\clearpage

\clearpage

\clearpage

\clearpage

\clearpage

\clearpage

\clearpage

\clearpage

\end{document}